\documentclass[iop]{emulateapj}   		

\usepackage{subfigure}					
\usepackage{graphicx}					
\usepackage{amssymb}					
\usepackage{mathtools}					
\usepackage{natbib}						

\usepackage{color}						
\usepackage{amsmath}
\usepackage{bm}						

\begin{document}

\title{Constraining AGN Feedback in Massive Ellipticals with South Pole Telescope Measurements of the Thermal Sunyaev-Zel'dovich Effect}
\author{Alexander Spacek, Evan Scannapieco, Seth Cohen, Bhavin Joshi, Philip Mauskopf}	
\affil{School of Earth and Space Exploration, Arizona State University,  PO Box 876004, Tempe - 85287, USA}


\begin{abstract}

Energetic feedback due to active galactic nuclei (AGNs) is likely to play an important role in the observed anti-hierarchical trend in the evolution of galaxies, and yet the energy injected into the circumgalactic medium by this process is largely unknown.  One promising approach to constrain this feedback is through measurements of spectral distortions in the cosmic microwave background due to the thermal Sunyaev-ZelÕdovich (tSZ) effect, whose magnitude is directly proportional to the energy input by AGNs.  With current instruments, making such measurements requires stacking large numbers of objects to increase signal-to-noise.   While one possible target for such stacks is AGNs themselves, these are relatively scarce sources that contain contaminating emission that complicates tSZ measurements.   Here we adopt an alternative approach and co-add South Pole Telescope SZ (SPT-SZ) survey data around a large set of massive quiescent  elliptical galaxies at $z \geq 0.5$, which are much more numerous and less contaminated than active AGNs, yet are subject to the same feedback processes from the AGNs they hosted in the past. We use data from the Blanco Cosmology Survey and VISTA Hemisphere Survey to create a large catalog of galaxies split up into two redshift bins: one with 3394 galaxies at $0.5\leq z \leq 1.0$ and one with 924 galaxies at $1.0\leq z \leq 1.5,$ with typical stellar masses of $1.5 \times 10^{11} M_\odot.$ We then co-add the emission around these galaxies, resulting in a measured tSZ signal at $2.2\sigma$ significance for the lower redshift bin and a contaminating signal at $1.1\sigma$ for the higher redshift bin. To remove contamination due to dust emission, we use SPT-SZ source counts to model a contaminant source population in both the SPT-SZ bands and \textit{Planck} high-frequency bands for a subset of 937 galaxies in the low-redshift bin and 240 galaxies in the high-redshift bin. This increases our detection to $3.6\sigma$ for low redshifts and $0.9\sigma$ for high redshifts. We find the mean angularly-integrated Compton-$y$ values to be $2.2_ {-0.7}^{+0.9} \times 10^{-7}$ Mpc$^2$ for low redshifts and $1.7_ {-1.8}^{+2.2} \times 10^{-7}$ Mpc$^2$ for high redshifts, corresponding to total thermal energies of $7.6_ {-2.3}^{+3.0} \times 10^{60}$ erg and $6.0_{-6.3}^{+7.7} \times 10^{60}$ erg, respectively. These numbers are higher than expected from simple theoretical models that do not include AGN feedback, and serve as constraints that can be applied to current simulations of massive galaxy formation. 

\end{abstract}

          

\section{Introduction}
\label{sec:intro}

In the prevailing model of galaxy formation, the collapse of baryonic matter follows the collapse of overdense regions of dark matter \citep[e.g.,][]{White1978,White1991,Kauffmann1993,Lacey1993}.  Over time, these dark matter halos accrete and merge to form deep gravitational potential wells.  These, in turn, lead to strong gravitationally powered shocks that cause the inflowing gas to be heated to high temperatures.   To collapse and form stars, the gas must radiate this energy away, a process that takes longer in the largest, most gravitationally bound structures \citep[e.g.,][]{Binney1977,Rees1977,Silk1977}.  Furthermore, galaxies also accrete and merge over time within their dark matter halos, a process that appears to be closely linked to the evolution of active galactic nuclei \citep[AGNs; e.g.,][]{Richstone1998,Cattaneo1999,Kauffmann2000}. Together, these processes point to a hierarchical picture in which larger star-forming galaxies, hosting larger AGNs,  form at later times as larger dark matter halos coalesce and more gas cools and condenses.

On the other hand, an increasing amount of observational evidence suggests that recent trends in galaxy and AGN evolution were anti-hierarchical. More massive galaxies appear to be forming stars at higher redshift, and since $z\approx2$ the characteristic mass of star-forming galaxies appears to have dropped by more than a factor of 3 \citep{Cowie1996,Brinchmann2004,Kodama2004,Bauer2005,Bundy2005,Feulner2005,Treu2005,Papovich2006,Noeske2007,Cowie2008,Drory2008,Vergani2008}. Similarly, since $z\approx2$ the characteristic AGN luminosity has dropped by more than a factor of 10, indicating that the typical masses of active supermassive black holes were larger in the past \citep{Pei1995,Ueda2003,Barger2005,Buchner2015}. While it has been argued that this observed ``downsizing'' is a natural result of the standard hierarchical framework \citep[e.g.,][]{Enoki2014}, most work has suggested that it requires additional heating of the circumgalactic medium by AGN feedback \citep{Merloni2004,Scannapieco2004,Scannapieco2005,Bower2006,Neistein2006,Thacker2006,Sijacki2007,Merloni2008,Chen2009,Hirschmann2012,Hirschmann2014,Mocz2013, Lapi2014, Schaye2015}. 

In a general AGN feedback model \citep[e.g.,][]{Scannapieco2005}, energetic AGN outflows due to broad absorption-line winds and/or radio jets blow cool gas out of the galaxy and/or heat the nearby intergalactic medium (IGM) enough to suppress the cooling needed to form further generations of stars and AGN. This quenching is redshift dependent, as the higher-redshift IGM is more dense and rapidly radiating and therefore a highly energetic outflow driven by a large AGN is required to have effective feedback. In the less dense lower-redshift IGM,  a less energetic outflow by a smaller AGN can produce similar cooling times. This means that at lower redshifts the AGNs in smaller galaxies can exert efficient feedback, preventing larger galaxies from forming stars, suppressing AGN accretion, and resulting in the cosmic downsizing that we observe.

There has been significant observational evidence of AGN feedback in action in galaxy clusters, primarily in the form of radio jets \citep{Schawinski2007,Rafferty2008,Fabian2012,Farrah2012,Page2012}. Galaxies near the center of clusters show a boosted likelihood of hosting large radio-loud jets of AGN-driven material \citep{Burns1990, Best2005, McNamara2005}, whose energies are comparable to those needed to stop the gas from cooling \citep[e.g.,][]{Simionescu2009}. Furthermore, AGN feedback from the central cD galaxies in clusters increases in proportion to the cooling luminosity, as expected in an operational feedback loop \citep[e.g.,][]{Birzan2004, Rafferty2006, Bruggen2009}.

Direct measurements of the characteristic heating of the interstellar medium (ISM) and surrounding IGM by AGN feedback have been more difficult due to the relatively high redshifts and faint signals involved. Broad absorption-line outflows (winds) are observed as blueshifted troughs in the rest-frame spectra of $\approx 20\%$ of all of quasars \citep{Hewett2003, Ganguly2008, Knigge2008}. However, quantifying AGN feedback requires estimating the mass-flux and the energy released by these outflows \citep[e.g.,][]{Wampler1995,deKool2001,Hamann2001,Feruglio2010,Sturm2011,Veilleux2013}. These quantities, in turn, can only be computed in cases for which it is possible to estimate the distance to the outflowing material from the central source, which is often highly uncertain.  While these measurements have been carried out for a select set of objects  \citep[e.g.,][]{Chartas2007,Moe2009,Dunn2010,Borguet2013,Chamberlain2015}, it is still unclear how these results generalize to AGNs as a whole.   At the same time it is still an open question whether AGN outflows triggered by galaxy interactions actually quench star formation in massive, high-redshift galaxies \citep[e.g.,][]{Fontanot2009,Pipino2009, Debuhr2010,Ostriker2010,Faucher2012,Newton2013,Feldmann2015}.

A promising method for quantifying the effect of AGN feedback is through measurements of the cosmic microwave background (CMB) radiation. The CMB has large-scale anisotropies that have been measured in great detail and provide insight into the cosmological parameters that shape our universe \citep[e.g.,][]{PlanckCollaboration2015}. At angular scales smaller than $\approx$ 5 arcmin, though, Silk damping washes out the primary CMB anisotropies \citep{Silk1968,PlanckCollaboration2015}, leaving room for secondary anisotropies. These include the Sunyaev-Zel'dovich effect, where CMB photons are scattered by hot, ionized gas \citep{Sunyaev1970,Sunyaev1972}. If the gas is sufficiently heated, inverse Compton scattering will shift the CMB photons to higher energies. This thermal Sunyaev-Zel'dovich (tSZ) effect directly depends on the thermal energy of the free electrons that the CMB radiation passes through, and it has a unique spectral signature that makes it well suited to measuring the heating of gas and characterizing AGN feedback \citep{Voit1994,Birkinshaw1999,Natarajan1999,Platania2002,Lapi2003,Chatterjee2007,Chatterjee2008,Scannapieco2008,Battaglia2010}. On the other hand, if an object is moving along the line of sight with respect to the CMB rest frame, then the Doppler effect will lead to an observed distortion of the CMB spectrum, referred to as the kinetic Sunyaev-Zel'dovich effect. The magnitude of this effect is proportional to the overall column depth of the gas times the velocity of the line of sight motion, and its spectral signature is indistinguishable from primary CMB anisotropies.

The expected tSZ distortion per source is too small to be detected by current instruments \citep[e.g.,][]{Scannapieco2008}, and so a stacking method must be applied to many sources in order to derive a significant signal from them. \citet{Chatterjee2009} found a tentative detection of quasar feedback using the Sloan Digital Sky Survey (SDSS) and Wilkinson Microwave Anisotropy Probe (WMAP), although it is ambiguous how much of their detected signal is due to AGN feedback and how much is due to other processes \citep[see][]{Ruan2015}. \cite{Hand2011} stacked $>$2300 SDSS-selected ``luminous red galaxies'' in data from the Atacama Cosmology Telescope (ACT) and found a $2.1\sigma-3.8\sigma$ tSZ detection after selecting radio-quiet galaxies and binning them by luminosity.  \cite{Gralla2014} stacked data from ACT at the positions of a large sample of radio AGN selected at 1.4 GHz to make a 5$\sigma$ detection of the tSZ effect associated with the haloes that host active AGN.  \citet{Greco2014} used Planck full mission temperature maps to examine the stacked tSZ signal of 188,042 ``locally brightest galaxies'' selected from the SDSS Data Release 7, finding a significant measurement of the stacked tSZ signal from galaxies with stellar masses above $\approx 2 \times 10^{11} M_\odot$.  \citet{Ruan2015} stacked Planck tSZ Compton-$y$ maps centered on the locations of 26,686 spectroscopic quasars identified from SDSS to estimate the mean thermal energies in gas surrounding such $z \approx 1.5$  quasars to be $\approx {10^{62}}$ erg. On the contrary, \citet{Cen2015} used a statistical analysis of stacked $y$ maps of quasar hosts using the \textit{Millennium Simulation} and found that, with the 10 arcmin full width at half maximum (FWHM) resolution of their Planck stacking process, the results of \citet{Ruan2015} could be explained by gravitational heating alone, with a maximum feedback energy of about 25\% of their stated value. In addition, they found that a 1 arcmin FWHM beam is much more favorable in distinguishing between quasar feedback models. \citet{Crichton2015} stacked $>$17,000 radio-quiet quasars from SDSS in ACT data and found $3\sigma$ evidence for the presence of associated thermalized gas and $4\sigma$ evidence for the thermal coupling of quasars to their surrounding medium. These initial tSZ AGN feedback measurements using quasars are promising, and they continue to motivate direct measurements that probe different AGN feedback regimes, especially at the 1 arcmin FWHM resolution of the South Pole Telescope (SPT) used in this work.

Although quasars are a popular target for measuring AGN feedback due to their brightness and their active feedback processes, their drawbacks are that they are relatively scarce and contain contaminating emission that obscures the signatures of AGN feedback. In this paper, we focus on measuring co-added tSZ distortions in the CMB around massive ($\geq10^{11} M_\odot$) quiescent elliptical galaxies at moderate redshifts ($0.5\leq z \leq 1.5$) using data from the Blanco Cosmology Survey \citep[BCS;][]{Desai2012}, VISTA Hemisphere Survey \citep[VHS;][]{Mcmahon2012}, and South Pole Telescope SZ survey \citep[SPT-SZ;][]{Schaffer2011}, in order to characterize the energy injected by the AGNs they once hosted. These galaxies contain almost no dust and are very numerous on the sky, making them well-suited for co-adding in large numbers in order to obtain good constraints on the energy stored in the gas that surrounds them.

The structure of this paper is as follows. In Section 2, we give an overview of the tSZ effect and provide a theoretical basis for our tSZ results. In Section 3, we describe the data that we use from the BCS, VHS, and SPT-SZ surveys. In Section 4, we describe our method of selecting optimal galaxies for our measurements. In Section 5, we describe how we generate a reliable catalog of sources and the parameters that describe their properties. In Section 6, we describe how we generate the final catalog of galaxies for our tSZ measurements. In Section 7, we describe our SPT-SZ filtering, the galaxy co-add process, and our overall results. This includes the initial measurements, $\chi^2$ statistics using just the SPT-SZ data, $\chi^2$ statistics incorporating Planck data, and a goodness-of-fit test using the Anderson-Darling (A-D) statistic. In Section 8, we summarize our results, discuss the implications for AGN feedback, and provide conclusions.

Throughout this work, we adopt a $\Lambda$ cold dark matter cosmological model with parameters \citep[from][]{PlanckCollaboration2015}, $h=0.68$, $\Omega_0$ = 0.31, $\Omega_\Lambda$ = 0.69, and $\Omega_b = 0.049$, where $h$ is the Hubble constant in units of 100 km s$^{-1}$ Mpc$^{-1}$, and $\Omega_0$, $\Omega_\Lambda$, and $\Omega_b$ are the total matter, vacuum, and baryonic densities, respectively, in
units of the critical density. All of our magnitudes are quoted in the AB magnitude system \citep[i.e.,][]{Oke1983}.



\section{Methods}
\label{sec:methods}


\subsection{The tSZ Effect}
\label{sec:tsz}

The tSZ effect describes the process by which CMB photons gain energy when passing through ionized gas \citep{Sunyaev1970,Sunyaev1972}. The photons are shifted to higher energies by thermally energetic electrons through inverse Compton scattering, and the resulting CMB anisotropy has a distinctive frequency dependence which causes a deficit of photons at frequencies below $\nu_{\text{null}} = 217.6 \, \text{GHz}$ and an excess of photons above $\nu_{\text{null}}$, with no change at $\nu_{\text{null}}$.  For the nonrelativistic plasma we will be interested in here, 
the change in CMB temperature as a function of frequency due to the tSZ effect is 
\begin{equation} 
\frac{\Delta T}{T_{\text{CMB}}} = y \left( x \frac{e^x + 1}{e^x - 1} - 4 \right),
\label{eq:DeltaT}
\end{equation}
where the  dimensionless Compton-$y$ parameter is defined as
\begin{equation}
 y \equiv \int dl \, \sigma_T \frac{n_e k \left( T_e - T_{\rm CMB} \right)}{m_e c^2}, 
 \label{eq:y}
\end{equation}
 where $\sigma_T$ is the Thomson cross section, $k$ is the Boltzmann constant, $m_e$ is the electron mass,
 $c$ is the speed of light, $n_e$ is the electron number density, $T_e$ is the electron temperature, $T_{\text{CMB}}$ is the CMB temperature (we use $T_{\text{CMB}} = 2.725$ K), and the integral is performed over the line of sight distance $l$. Finally, the dimensionless frequency $x$ is given by
\begin{equation} 
x \equiv \frac{h \nu}{k T_{\text{CMB}}} = \frac{\nu}{56.81 \, \text{GHz}} ,
\label{eq:x}
\end{equation}
where $h$ is the Planck constant.

We can calculate the total excess thermal energy associated with a source by integrating Equation (\ref{eq:y}) over a region of sky around the source as \citep[e.g.,][]{Scannapieco2008,Ruan2015},
\begin{equation}
\begin{split}
\int d\bm{\theta} \, y(\bm{\theta})   &= \int d\bm{\theta} \int dl \, \sigma_T \frac{n_e k  T_e }{m_e c^2} \\ &= \frac{\sigma_T}{m_e c^2} l_{\rm ang}^{-2} \int dV \, n_e k T_e   ,
\label{eq:thetay}
\end{split}
\end{equation}
where $\bm{\theta}$ is a two-dimensional vector in the plane of the sky in units of radians, $l_{\rm ang}$ is the angular diameter distance to the source, $V$ is the volume of interest around the source,
and we have restricted our attention to hot gas with $T_e \gg T_{\rm CMB}.$
In Equation (\ref{eq:thetay}), the Compton-$y$ integral has become a volume integral of the electron pressure (i.e. $P_e = n_e k T_e$), which is related to the associated thermal energy as
\begin{equation} 
\int dV \, n_e k T_e = \left( \frac{2}{3} \right) \left( \frac{1+A}{2+A} \right) E_{\rm therm},
\label{eq:Pe}
\end{equation}
where $A = 0.08$ is the cosmological number abundance of helium, and
$E_{\rm therm}$ is the total  thermal energy associated with the source: that gained from the initial collapse of the baryons, plus the contribution from the AGN, minus the losses due to cooling and the $P dV$ work done during expansion.
We can combine Equations (\ref{eq:thetay}) and (\ref{eq:Pe}) and solve for $E_{\rm therm}$ to get
\begin{equation} 
\begin{split}
E_{\rm therm} &= 2.9\frac{m_e c^2}{\sigma_T} l_{\rm ang}^2  \int d\bm{\theta} y(\bm{\theta}) \\
&= 2.9 \times 10^{60} {\rm erg} \, \left(\frac{l_{\rm ang}}{\text{Gpc}}\right)^2 
\frac{\int d\bm{\theta} y(\bm{\theta})}{10^{-6} \, \text{arcmin$^2$}}.
\label{eq:Ethrm}
\end{split}
\end{equation}
Finally, we can combine Equations (\ref{eq:DeltaT}) and (\ref{eq:Ethrm}) to get $E_{\rm therm}$ in terms of $\Delta T$ at a given dimensionless frequency $x$,
\begin{equation}
E_{\rm therm} = \frac{1.1 \times 10^{60} {\rm erg}}{x \frac{e^x + 1}{e^x - 1} - 4} \, \left(\frac{l_{\rm ang}}{\text{Gpc}}\right)^2 
\frac{\int \Delta T(\bm{\theta}) d\bm{\theta}}{\text{$\mu$K arcmin$^2$}}.
\label{eq:EthrmT}
\end{equation}



\pagebreak

\subsection{Models of Gas Heating}
\label{sec:gas}

To compare the energies and angular sizes above with the expectations from models of feedback, we can construct a simple model of gas heating with and without AGN feedback.   
To do this we first compute $R_{\rm vir},$ the virial radius of a (spherical) dark matter halo defined as the physical radius within which the density is 200 times the mean cosmic value.  As a function of redshift $z$ and mass $M,$ this is
\begin{equation}
\begin{split}
R_{\rm vir} &= \left[\frac{M}{(4\pi/3) 200 \Omega_0 \rho_{\rm crit} (1+z)^3} \right]^{1/3} \\ &=  0.67 \, {\rm Mpc} \, M_{13}^{1/3} (1+z)^{-1},
\label{eq:Rvir}
\end{split}
\end{equation}
where $\rho_{\rm crit}$ is the critical density at $z=0$, and  $M_{13}$ is the mass of the halo in units of $10^{13} M_\odot.$
This can be compared to the angular scales above, using the fact that  at an angular diameter distance of 1 Gpc, 1 arcminute corresponds to 0.29 Mpc.

If the gas collapses and virializes along with the dark matter, it will be shock-heated during gravitational infall to the virial temperature, 
\begin{equation} 
T_{\rm vir} = \frac{G M}{R_{\rm vir}}   \frac{\mu m_p}{2 k}  = 2.4 \times 10^6 {\rm K} \, M^{2/3}_{13}  (1+z),
\label{model:Tvir}
\end{equation}
where $G$ is the gravitational constant, $m_p$ is the proton mass, and $\mu = 0.62$ is the average particle mass in units of $m_p.$
If we approximate the gas distribution as isothermal at this temperature, its
total thermal energy can then be estimated as 
\begin{equation}
\begin{split}
E_{\rm therm,gravity} &= \frac{3 k T_{\rm vir}}{2} \frac{\Omega_b}{\Omega_0} \frac{M}{\mu m_p} \\  &=
1.5 \times 10^{60} \, {\rm erg} \, M_{13}^{5/3} (1+z).
\label{eq:Etherm}
\end{split}
\end{equation}

To relate the stellar masses of the galaxies we will be stacking to the dark matter halo masses, we can take advantage of 
the observed relation between black hole mass and halo circular velocity, $v_c,$ 
from \citet[][see also \citealp{Merritt2001,Tremaine2002}]{Ferrarese2002}, and convert the black hole mass to its corresponding bulge dynamical mass using a factor of 400 \citep{Marconi2003}.
This  gives
\begin{equation}
\begin{split}
M_{\rm stellar} &= 6.6_{-3.2}^{+5.5}  \times 10^{10} \, M_\odot \,\left(\frac{v_{\rm c}}{300 \, {\rm km~s^{-1}}}\right)^{5}\\
 &= 2.8_{-1.4}^{+2.4}  \times 10^{10} M_\odot   \,  M_{13}^{5/3} (1+z)^{5/2},
 \label{eq:Mstellar}
\end{split}
\end{equation}
where we have used the fact that $v_c = (G M/R_{\rm vir})^{1/2} = 254 \, {\rm km \, s}^{-1} \, M_{13}^{1/3} (1+z)^{1/2},$
and taken $M_{\rm stellar} \propto v_c^{\alpha_c}$ with the power law index $\alpha_c=5,$ which is near the center of the allowed range of $5.4 \pm 1.1, $ and we take our uncertainties from \citet{Ferrarese2002}.
Substituting Equation (\ref{eq:Mstellar}) into Equation (\ref{eq:Etherm}) gives
\begin{equation}
\begin{split}
E_{\rm therm,gravity}  &= 5.4_{-2.9}^{+5.4} \times 10^{60} \, {\rm erg} \\ &\times \frac{M_{\rm stellar}}{10^{11} M_\odot} (1+z)^{-3/2}.
\label{eq:Egrav}
\end{split}
\end{equation}
This is the total thermal energy expected around a galaxy of stellar mass $M_{\rm stellar}$ due  purely to gravitational heating,
and ignoring both radiative cooling, which will decrease $E_{\rm therm},$ and AGN feedback, which will increase it.

While there are many models of AGN feedback, each of which will lead to somewhat different signatures in our data, we can estimate the overall magnitude of this effect by making use of the simple model described in \citet[][see also \citealp{Thacker2006,Scannapieco2008}]{Scannapieco2004}.
In this case, AGN feedback is described as tapping into a small fraction, $\epsilon_{k},$ of the total bolometric luminosity of the AGN to heat the surrounding gas.   In particular, black holes are assumed to shine at the Eddington luminosity ($1.2 \times 10^{38}$ erg ${\rm s}^{-1}$ $M_\odot^{-1}$ ) for a time $0.035 \, t_{\rm dynamical}$,
where
\begin{equation}
t_{\rm dynamical} \equiv R_{\rm vir}/v_c = 2.6 \, {\rm Gyr} \, (1+z)^{-3/2}.
\label{eq:tdyn}
\end{equation}
This choice of timescale gives a good match to the observed evolution of the quasar luminosity function \citep{Wyithe2002,Wyithe2003,Scannapieco2004}.
This gives
\begin{equation}
\begin{split}
E_{\rm therm, feedback}  
 		&=   4.1 \times 10^{60} \, {\rm ergs} \,  \epsilon_{k,0.05} \\ &\times  \frac{M_{\rm stellar}}{10^{11} M_\odot} \, (1+z)^{-3/2}.
\label{eq:EAGN}
\end{split}
\end{equation}
Here $\epsilon_{k,0.05} \equiv \epsilon_{k}/0.05$, such that the kinetic energy input is normalized to a typical value needed to achieve antiheirarchical galaxy evolution through effective feedback \citep[e.g.,][]{Scannapieco2004,Thacker2006,Costa2014}. Note that the uncertainty in this equation is completely dominated by the $\epsilon_{k,0.05}$ term, which is uncertain to within an order of magnitude.

This energy input is equal in magnitude to  the errors in $E_{\rm therm,gravity},$ meaning that the differences between models with and without AGN feedback will not be dramatic.   Thus only detailed simulations will be able to make precise predictions on the level needed to rule out or lend support to a particular model of AGN feedback. Although carrying out such simulations is beyond the scope of this paper, Equations (\ref{eq:Egrav}) and (\ref{eq:EAGN}) are roughly consistent with such sophisticated models \citep[e.g.,][]{Thacker2006,Chatterjee2008}, meaning that they can be used as an approximate guide to interpreting our results. Thus, we will use them to provide a general context for thinking about our observational results in terms of AGN feedback.

Finally, we note that the sound speed $c_s$ of the gas in the gravitationally heated case is similar to the circular velocity (i.e.,  $c_s = [\gamma kT/(\mu m_p)]^{1/2} = (\gamma/2)^{1/2} v_c$, where $\gamma$ is the adiabatic index), and the expected energy input from the AGN is similar to the energy input from gravitational heating.  This means that  the energy input from the AGN will take a timescale $\approx t_{\rm dynamical}$ to impact gas on the scale of the halo, and it is unlikely to affect scales much larger than $\approx 2 R_{\rm vir}.$   These sizes and timescales mean that at the moderate redshifts we will be exploring, the majority of the gas heating we are interested in will occur on scales $\lesssim 2$ arcmin.    


\section{Data}
\label{sec:data}

\begin{figure}[t]
\centerline{\includegraphics[height=8cm]{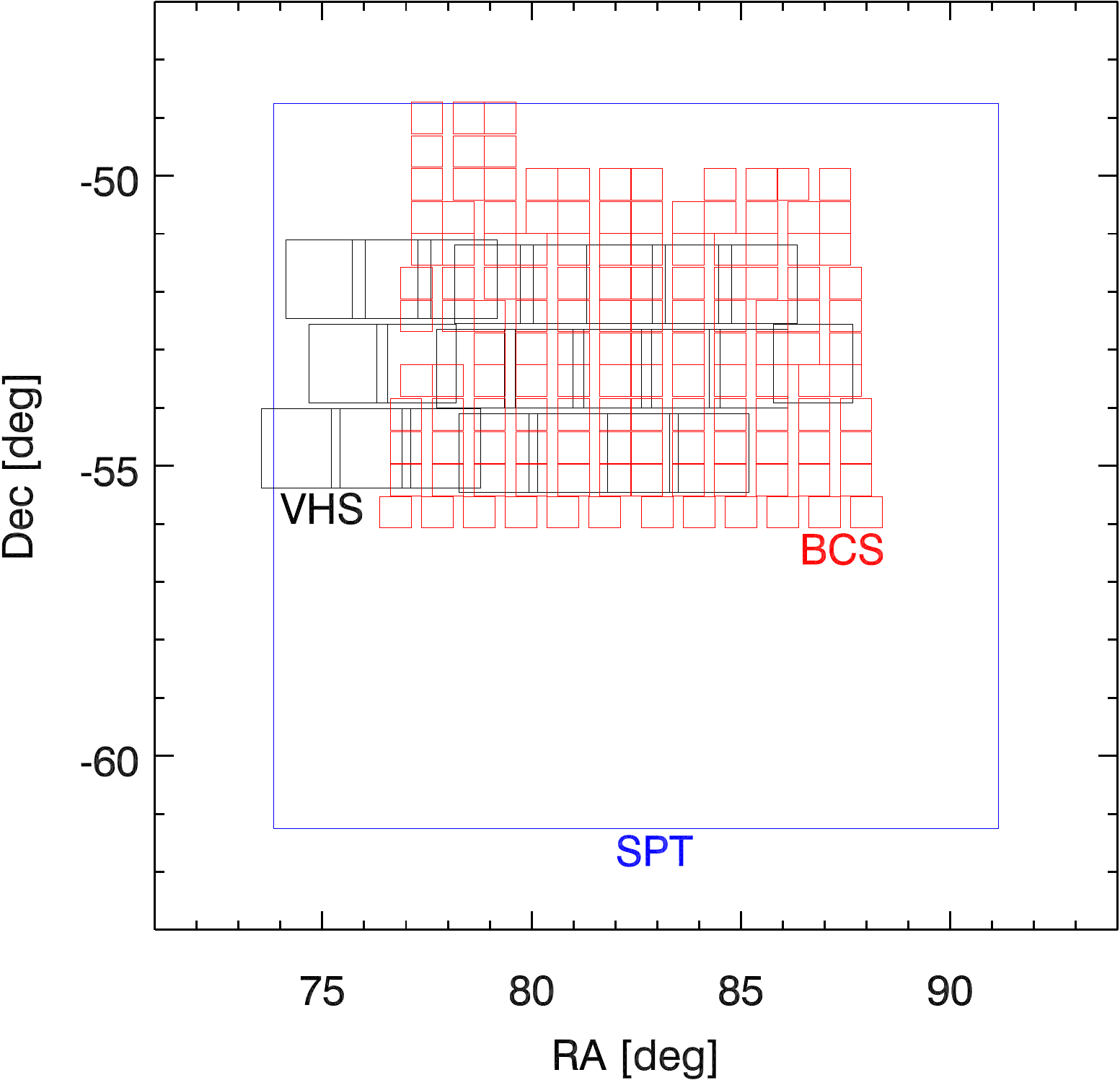}}
\caption{\small Approximate locations on the sky for the overlapping BCS tiles (\textit{red}), VHS tiles (\textit{black}), and SPT-SZ field (\textit{blue}). \vspace{4mm}}
\label{fig:tiles}
\end{figure}

Three public datasets are critical to our analysis. To detect, measure, and select galaxies, we use optical and infrared data from the BCS \citep{Desai2012} and infrared data from the VHS \citep{Mcmahon2012}. To stack microwave  observations to detect the tSZ signal, we use data from the SPT-SZ survey \citep{Schaffer2011}. The three datasets overlap over an area of $\approx$ 43 deg$^2$, as can be seen in Figure \ref{fig:tiles}, and provide good wavelength coverage and sensitivities, as can be seen in Table \ref{tab:bands}.  Here we describe each of these data sets in turn.

\begin{table*}[t]
\begin{center}
\resizebox{12cm}{!}{
\begin{tabular}{|c|c|c|c|c|c|}
	\hline
	Filter & Center [nm] & Width [nm] & Depth [AB] & Seeing [FWHM] & Survey \\ \hline
	$g$ & 481.3 & 153.7 & 23.9 & 1.0 arcsec & BCS\footnotemark \\
	$r$ & 628.7 & 146.8 & 24.0 & 1.0 arcsec & BCS\footnotemark[1] \\
	$i$ & 773.2 & 154.8 & 23.6 & 0.8 arcsec & BCS\footnotemark[1] \\
	$z$ & 940.0 & 200.0 & 22.1 & 0.9 arcsec & BCS\footnotemark[1] \\
	$J$ & 1252 & 172.0 & 20.86 & 1.1 arcsec & VHS DES\footnotemark \\
	$H$ & 1645 & 291.0 & 20.40 & 1.0 arcsec & VHS DES\footnotemark[2] \\
	$K_s$ & 2147 & 309.0 & 20.16 & 1.0 arcsec & VHS DES\footnotemark[2] \\ \hline
	$150 \text{GHz}$ & 153.4 GHz & 35.2 GHz & 17 $\mu$K-arcmin & 1.15 arcmin & SPT-SZ\footnotemark \\
	$220 \text{GHz}$ & 219.8 GHz & 43.7 GHz & 41 $\mu$K-arcmin & 1.05 arcmin & SPT-SZ\footnotemark[3] \\ \hline
\end{tabular}
}
\end{center}
\vspace{-2mm}
\footnotesize{[a] http://www.ctio.noao.edu/noao/content/MOSAIC-Filters; \citet{Desai2012}}\\
\footnotesize{[b] http://casu.ast.cam.ac.uk/surveys-projects/vista/technical/filter-set; \citet{Mcmahon2012}}\\
\footnotesize{[c] \citet{Schaffer2011}}
\vspace{-2mm}
\caption{\small Band/filter information. BCS depths are 10$\sigma$ AB magnitude point source depths; VHS depths are 5$\sigma$ median AB magnitude depths; SPT depths use a Gaussian approximation for the beam.}
\label{tab:bands}
\vspace{4mm}
\end{table*}

\subsection{The BCS}
\label{sec:bcs}

The BCS was a National Optical Astronomy Observatory (NOAO) Large Survey project that observed $\approx$ 80 deg$^2$ of the southern sky over 60 nights between 2005 November and 2008 November on the 4m V\'ictor M. Blanco telescope at the Cerro Tololo Inter-American Observatory in Chile using the Mosaic II imager with $g$, $r$, $i$, and $z$ bands \citep{Desai2012}. The filter centers, effective widths, and magnitude limits are given in Table \ref{tab:bands}.

The BCS data is split up into many smaller 36 $\times$ 36 arcmin (8192 $\times$ 8192 pixel) images called tiles, with $\approx$1 arcmin overlap between neighboring tiles. Each pixel subtends 0.27 arcsec on the sky. As described in \citet{Desai2012}, each raw data tile is put through a detrending pipeline, which consists of crosstalk corrections, overscan, flatfield, bias and illumination correction, and astrometric calibration. The average FWHM of the seeing disk in the single epoch images ranges between 0.7 and 1.6 arcsec. Each tile is then put through a co-addition pipeline that combines data taken over the same locations on the sky to build deeper single images. This results in a co-added tile image and an inverse-variance weightmap (confidence image) for each tile region of the survey. We use the area of the BCS that overlaps with the VHS and SPT-SZ, known as the \textit{5 hr field} (referring to its right ascension). This dataset consists of 135 tiles and their associated weightmaps for each band, covering $\approx$ 45 deg$^2$.


\subsection{The VHS}
\label{sec:vhs}

The VHS is a large-scale near-infrared survey  whose goal is to survey the entire southern celestial hemisphere ($\approx$ 20,000 deg$^2$; \citealp{Mcmahon2012}). The survey component in which we are interested is called the VHS DES (DES because it overlaps with the Dark Energy Survey), a 5000 deg$^2$ region that is imaged with 120 s exposure times in the $J$, $H$, and $K_s$ bands (see Table \ref{tab:bands}). The data was obtained from 2009 to 2011 on the 4.1 m Visible and Infrared Survey Telescope for Astronomy (VISTA) at the Paranal Observatory in Chile.

The VHS data is split up into $\approx$ 2 $\times$ 1.5 deg ($\approx$12,770 $\times$ 15,660 pixel) tiles. Each pixel subtends 0.33 arcsec on the sky. As described on the VISTA data processing web page\footnote{\scalebox{0.8}{http://casu.ast.cam.ac.uk/surveys-projects/vista/technical/data-processing}}, the raw VHS data go through a pipeline that involves reset correction, dark correction, linearity correction, flat field correction, sky background correction, destripe, jitter stacking, astrometric and photometric calibration, and tile generation. Tiles are generated from six smaller, stacked pawprints, each containing 16 even smaller detector-level images\footnote{\scalebox{0.8}{http://casu.ast.cam.ac.uk/surveys-projects/vista/technical/tiles}}, and the median image seeing as measured from stellar FWHM on VHS pawprints ranges from 0.89 arcsec in $K_s$ to 0.99 arcsec in $J$.  The stacked paw prints then  result in a science-ready tile image and inverse-variance weightmap for each tile region of the survey. We are interested in the area of the VHS that overlaps with the BCS (see Figure 1). This results in 20 tiles and their associated weightmaps, covering $\approx$ 55 deg$^2$.


\subsection{The SPT-SZ Survey}
\label{sec:spt}

The SPT-SZ survey \citep{Schaffer2011} used the 10m South Pole Telescope (SPT) at the National Science Foundation's (NSF) Amundsen-Scott South Pole Station  to survey a large area of the sky at millimeter and sub-millimeter wavelengths with arcminute angular resolution and low noise \citep{Ruhl2004, Padin2008, Carlstrom2011}. The  survey observed 2500 deg$^2$ of the southern sky during the austral winter seasons of 2008 through 2011. Data from the 2011 release that we are using covers $\approx$ 95 deg$^2$ to a depth of 17 and 41 $\mu$K arcmin at 150 GHz and 220 GHz, respectively, centered at (R.A., decl.) = (82.7, -55) degrees (see Table \ref{tab:bands}).

The SPT-SZ data is contained in a single image per band, $\approx 20^\circ \times 10^\circ$ ($\approx$3000 $\times$ 3000 pixels) projected as either a Sanson-Flamsteed projection or an oblique Lambert equal-area azimuthal projection. The Sanson-Flamsteed projection is most useful for cluster-finding and contains masked point-sources, while the Lambert projection is most useful for point-source analysis. Since we are interested in individual sources that are undetected and within the noise level, we use the Sanson-Flamsteed projection with point-sources masked. Each pixel subtends 15 arcsec on the sky. As described in \citet{Schaffer2011}, the raw data goes through a pre-processing stage where the data from a single observation of the field is calibrated, data selection cuts are applied, and initial filtering and instrument characterization are performed. A map-making stage with additional filtering is performed on the pre-processed data and the data are binned into single-observation maps used for final co-adds. The final data products include a co-added image, two-dimensional beam functions, filter transfer functions, and noise power spectral densities for each band.

It is worth noting that 220 GHz is very close to the frequency at which there is no change in the CMB due to the tSZ effect ($\nu_{\text{null}}$ = 217.6 GHz), while 150 GHz, which is close to the peak of the undistorted CMB spectrum (160 GHz), will see a decrement in radiation. Equations (\ref{eq:DeltaT}) and (\ref{eq:x}) can be rewritten for these bands, though the equations must now involve integration over the SPT filter curves. Once this is done, we can write the Compton-$y$ parameter as
\begin{equation} 
y = -0.41\,\frac{\Delta T_{150}}{1 \text{K}} \hspace{1cm} {\rm and}  \hspace{1cm} y = 9.9\,\frac{\Delta T_{220}}{1 \text{K}},
\label{eq:newys}
\end{equation}
where $\Delta T_{150}$ and $\Delta T_{220}$ are the temperature anisotropies at 150 and 220 GHz, respectively.
Here we can explicitly see that, for the same $y$, the increase in $\Delta T_{220}$ is about 24 times less than the decrease in $\Delta T_{150}$. A measurement of the tSZ effect is therefore expected to give us a clear decrement at 150 GHz and no detectable change at 220 GHz.

We can also use Equations (\ref{eq:x}) and (\ref{eq:EthrmT}), integrated over the SPT filter curve as mentioned above, to compare the tSZ decrement at 150 GHz to the total thermal energy of an object as
\begin{equation} 
E_{\rm therm} = -1.2\times 10^{60} \text{ergs} \left(\frac{l_{\rm ang}}{\text{Gpc}}\right)^2 \frac{\int \Delta T_{150}(\bm{\theta}) d\bm{\theta}}{\text{$\mu$K arcmin$^2$}}.
\label{eq:newEthrm}
\end{equation}
Given the arcminute angular resolution and 17 $\mu$K arcmin sensitivity of the SPT 150 GHz data, this means that for stacks of several thousand sources we can hope to derive constraints on the order of $\Delta E_{\rm therm} \approx 10^{60}$ erg. This is sufficient to derive constraints that are interesting for discriminating between models of AGN feedback, as discussed in Section \ref{sec:gas}.


\section{Selecting Galaxies to Constrain AGN Feedback}
\label{sec:galselect}

\begin{figure*}[ht]
\centerline{
\includegraphics[height=8.9cm]{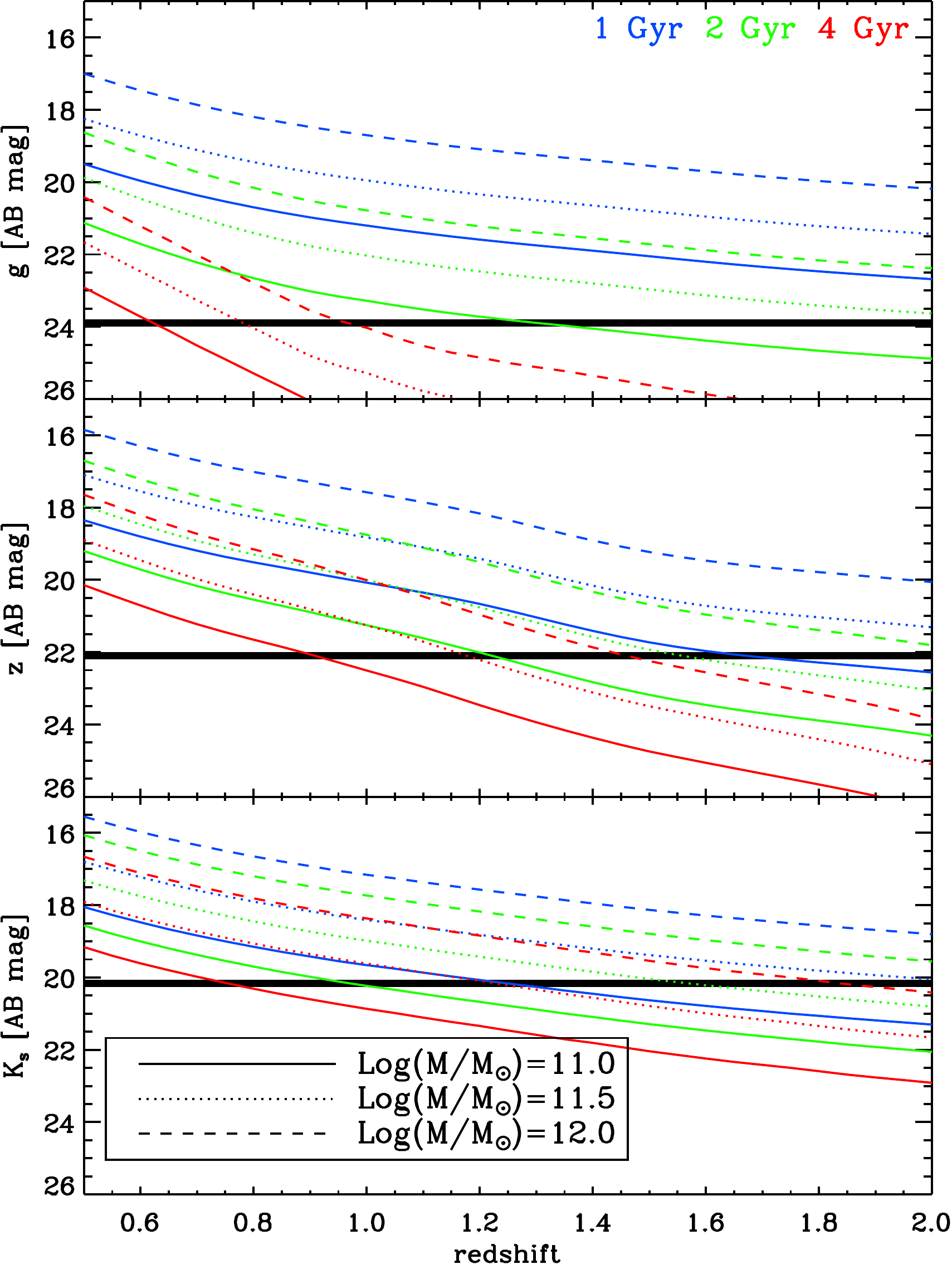} \qquad \qquad
\includegraphics[height=8.9cm]{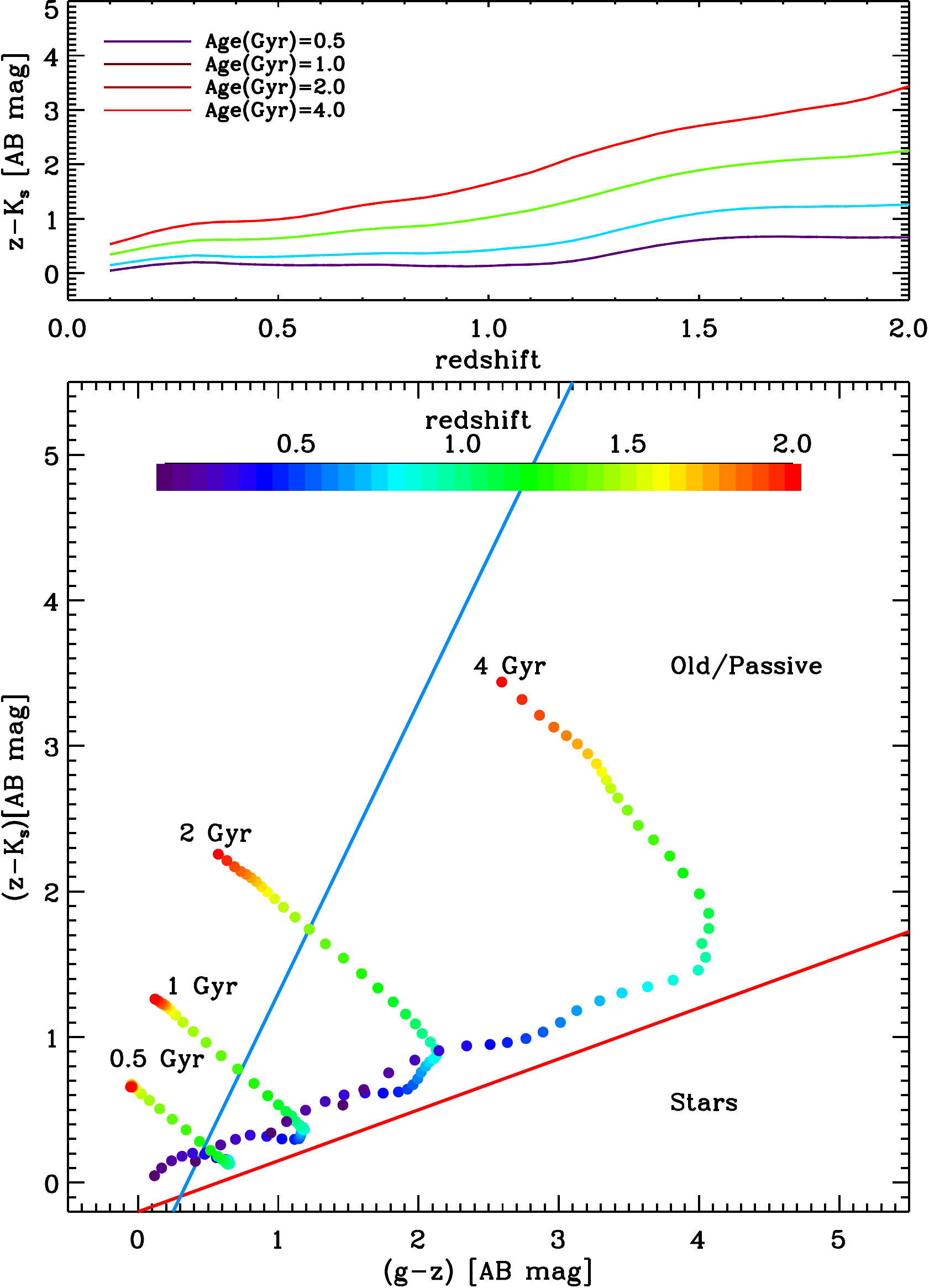}}
\caption{\small \textit{Left}: Optical and infrared magnitudes of early-type galaxies as a function of mass (indicated by line type), age (indicated by color), and redshift, as compared with limits from VHS ($K_s$) and BCS ($g$, $z$; solid black lines). \textit{Upper-right}: Color-redshift plot showing how age and redshift are distinguished in a galaxy's $z-K_s$ color. \textit{Lower-right}: Color-color plot illustrating how passive $z \geq 0.5$ galaxies are easily distinguished from stars and young galaxies. The red and blue lines represent Equations (\ref{eq:stars}) and (\ref{eq:passivecut}), respectively. \vspace{5mm}}
\label{fig:candidates}
\end{figure*}

If we compare Equation (\ref{eq:Ethrm}) with (\ref{eq:Egrav}) and (\ref{eq:EAGN}), we can see that achieving constraints on $E_{\rm therm}$ at the level to discriminate between the models above requires measurements with sensitivities on the order of  $\int d\bm{\theta} y(\bm{\theta}) \approx 10^{-6} \, {\rm arcmin}^2.$ With current instruments at arcminute resolution, this requires stacking $\gtrsim1000$ sources.  Thus, the  first step to constraining AGN feedback is selecting an appropriate set of objects around which to co-add CMB data.  Here one must balance several competing concerns. At the earliest times, when the most luminous AGN are in the midst of heating the surrounding gas, tSZ measurements are particularly difficult. This is both because emission from the AGN and its host are likely to contribute in the interesting 100-300 GHz frequency range, as well as because the low number density of such sources makes it very difficult to co-add them in meaningful numbers using the SPT data we are working with \citep[although with a large field and spectral energy distribution (SED) fitting it is possible to extract a tSZ signal, i.e.][]{Gralla2014}. On the other hand, the cooling times of regions heated by the most luminous AGNs are likely to be longer than the Hubble time \citep[e.g.,][]{Scannapieco2004}, making the heated gas we are interested in detectable long after the active AGN phase has passed. Furthermore, at the lowest redshifts, the largest bulge galaxies will be absorbed into galaxy clusters, where gravitational heating effects will be sufficiently large as to make AGN feedback processes difficult to distinguish. 

For these reasons, we restrict our attention to elliptical galaxies, rather than luminous AGNs or dusty late-type galaxies, and select only galaxies with redshifts greater than $z=0.5.$ The left panel of Figure \ref{fig:candidates} illustrates the $g,$ $z,$ and $K_s$ band magnitudes of ellipticals as a function of age and mass, computed from GALAXEV population synthesis models (\citealp{Bruzual2003}; for band information see Table \ref{tab:bands}). Here we have taken a star formation history $\propto$ exp(-t/$\tau$), where $\tau$ = 0.51 Gyr, and ages of 1, 2, and 4 Gyr.  Note that in the standard cosmology, the ages of the universe at $z$ = 0.5, 1, 1.5, and 2.0 are 8.5, 5.9, 4.3, and 3.3 Gyr, respectively.

In Figure \ref{fig:candidates} we have also plotted the magnitude limits of the BCS and VHS data.   By comparing the models and limits we can see that large passive galaxies are indeed detectable in this data at a wide range of  redshifts above $z=0.5$.  In particular, galaxies with ages $\approx$ 1 Gyr with stellar masses above $10^{11} \, M_\odot$  are detectable in both the optical and infrared data from $z=0.5$ to 1.2 while 1 Gyr galaxies with stellar masses above $10^{11.5} \, M_\odot$ are detectable out to $z=2$.  At ages of 2 Gyr,  galaxies with stellar masses above 10$^{11}$ $M_\odot$ are detectable out to $z=1.0$  and  galaxies with stellar masses above 10$^{11.5}$ $M_\odot$ are detectable out to $z=1.6$.  Finally, for an age of 4 Gyr, galaxies with stellar masses above $10^{11} \, M_\odot$ can be detected out to  $z=0.7,$ and galaxies with stellar masses above $10^{11.5} \, M_\odot$ can be detected out to $z=1.2$.

The right panel of Figure \ref{fig:candidates} shows that we can also use $g-z$ vs. $z-K_s$ colors to cleanly separate $\gtrsim 1$ Gyr old galaxies at $0.5\leq z \leq 1.5$ from stars and star-forming systems, making use  of the $gzK_s$ method outlined in \citet{Arcila-Osejo2013} \citep[see also][]{Daddi2004,Cameron2011}.
In particular, by applying a cut
\begin{equation} 
(z - K_s) \geq 0.35 (g - z) - 0.2,
\label{eq:stars}
\end{equation}
where $g,$ $z,$ and $K_s$ are AB magnitudes, we are able to separate the galaxies we are interested in from Galactic stars. 
Furthermore, by applying a cut 
\begin{equation} 
(z-K_s)  \leq 2 (g-z) - 0.7,
\label{eq:passivecut}
\end{equation}
we can also separate  passively evolving galaxies from young galaxies over the full redshift range in which we are interested.   Taken together, these results make clear that if we focus on the redshift range $0.5-1.5,$ we should be able to efficiently select a large number of suitable galaxies from the BCS and VHS data that we are using.


\section{Creating a Catalog of Galaxies}
\label{sec:galcat}


\subsection{Image Matching}
\label{sec:matching}

As seen in  Figure \ref{fig:candidates},  the sources we are interested in are brightest in the $K_s$ band, and thus we use it to make all of our detections. Because  the BCS and VHS tiles are different sizes and in different locations (see Figure \ref{fig:tiles}), we consider every possible overlap between images when aligning the other data to the $K_s$ tiles.   We then match pixel sizes and locations, and to insure that fixed aperture flux measurements are consistent between bands, we also match the seeing between the $K_s$ tiles and the other bands.

If the $K_s$ tile has worse seeing than the other band, we simply degrade the other image with a Gaussian filter until it matches
the FWHM of the $K_s$ image.  On the other hand,  if the $K_s$ image has better seeing, we degrade it to match the other band, compute the ratio of 3 arcsec diameter aperture fluxes between the two bands described below, and finally multiply the ratio by the  3 arcsec diameter flux measured from the unconvolved $K_s.$  That is, we compute and apply an aperture correction as  Flux$_{grizJH}$ = Flux$_{grizJH,0}$ $\times$ (Flux$_{K_s,0}$ / Flux$_{K_s,\text{degraded}}$), where $0$ denotes the non-degraded measurement.  This is done to preserve both accurate colors and the best possible $K_s$ flux in every case, since that is the most important band for our purposes. Note that, since the seeing is nearly the same for all tiles ($\approx 1$ arcsec, see Table \ref{tab:bands}), this correction is minor, with a mean ratio (Flux$_{K_s,0}$ / Flux$_{K_s,\text{degraded}}$) of 1.026 across all tiles.

\begin{table*}[t]
\begin{center}
\resizebox{14cm}{!}{
\begin{tabular}{|c|c|c|c|}
	\hline
	Configuration parameter & Value & Configuration parameter & Value \\ \hline
	\texttt{DETECT\_TYPE} & ccd & \texttt{PHOT\_AUTOPARAMS} & 2.5,3.5 \\
	\texttt{DETECT\_MINAREA} & 4 & \texttt{PHOT\_AUTOAPERS} & 0.0,0.0 \\
	\texttt{THRESH\_TYPE} & relative & \texttt{SATUR\_LEVEL} & 32,000 (VHS) \\
	\texttt{DETECT\_THRESH} & 3.0 & $\cdot\cdot\cdot$ & 20,000 (BCS) \\
	\texttt{ANALYSIS\_THRESH} & 3.0 & \texttt{GAIN} & 4.2 (VHS)  \\
	\texttt{FILTER} & y & $\cdot\cdot\cdot$ & 0 (BCS) \\
	\texttt{FILTER\_NAME} & gauss\_3.0\_3x3.conv & \texttt{PIXEL\_SCALE} & 0 \\
	\texttt{DEBLEND\_NTHRESH} & 32 & \texttt{BACKPHOTO\_TYPE} & local  \\
	\texttt{DEBLEND\_MINCONT} & 0.005 & \texttt{BACKPHOTO\_THICK} & 24  \\
	\texttt{CLEAN} & y & \texttt{BACK\_TYPE} & auto  \\
	\texttt{CLEAN\_PARAM} & 1.0 & \texttt{BACK\_VALUE} & 3.0  \\
	\texttt{MASK\_TYPE} & correct & \texttt{BACK\_SIZE} & 64 \\
	\texttt{WEIGHT\_TYPE} & map\_weight & \texttt{BACK\_FILTERSIZE} & 3 \\
	\texttt{WEIGHT\_GAIN} & n & \texttt{BACK\_FILTERTHRESH} & 0.0  \\
	\texttt{RESCALE\_WEIGHTS} & y & \texttt{MEMORY\_OBJSTACK} & 10,000 \\
	\texttt{PHOT\_APERTURES} & 9 & \texttt{MEMORY\_PIXSTACK} & 1,500,000 \\
\hline
\end{tabular} }
\end{center}
\vspace{-2mm}
\caption{\small SExtractor input parameters for all $K_s$-aligned tiles.}
\label{tab:sex}	
\end{table*}

\subsection{Detecting and Measuring Sources}
\label{sec:sex}

To detect and measure every object in our field, we use the SExtractor software package, version 2.8.6 \citep{Bertin1996}\footnote{SExtractor v2.13 User's manual, E. Bertin}$^{, }$\footnote{http://www.astromatic.net/software/sextractor}. The code detects and measures sources in an image through the following five-step process:
(i) it creates a background map that estimates the noise at every pixel in the image; (ii)  it detects sources using a thresholding technique; (iii) it uses a multiple isophotal analysis technique to deblend objects; (iv) it throws out spurious detections made in the wings of larger objects; and (v) it estimates the flux of each remaining object.  Each of these steps can be adjusted by the user through configuration parameters, and we list our choice of these parameters for both BCS and VHS tiles in Table \ref{tab:sex}.

In all cases, we use SExtractor's dual-image mode, which allows us to make flux measurements in all bands from the same sources detections in the $K_s$ band. This results in a catalog of $K_s$-detected sources with \texttt{MAG\_AUTO} and 3 arcsec diameter aperture flux measurements in every band.  We use corrected  \texttt{MAG\_AUTO} for our final catalogs and the 3 arcsec diameter aperture  \texttt{MAG\_APER} to compute aperture corrections as described in Section \ref{sec:matching}. Finally, the overlap between tiles within both the BCS and VHS images leads to some sources being detected in multiple tiles.  To correct for this, we match our catalog with itself and  remove multiple occurrences of sources within 1 arcsec of each other. At this point in the analysis, our full catalog contains 565,561 sources, 168,944 (30\%) of which are identified as duplicates. This leaves 396,617 total sources.

\begin{figure}[h]
\centerline{\includegraphics[height=8cm]{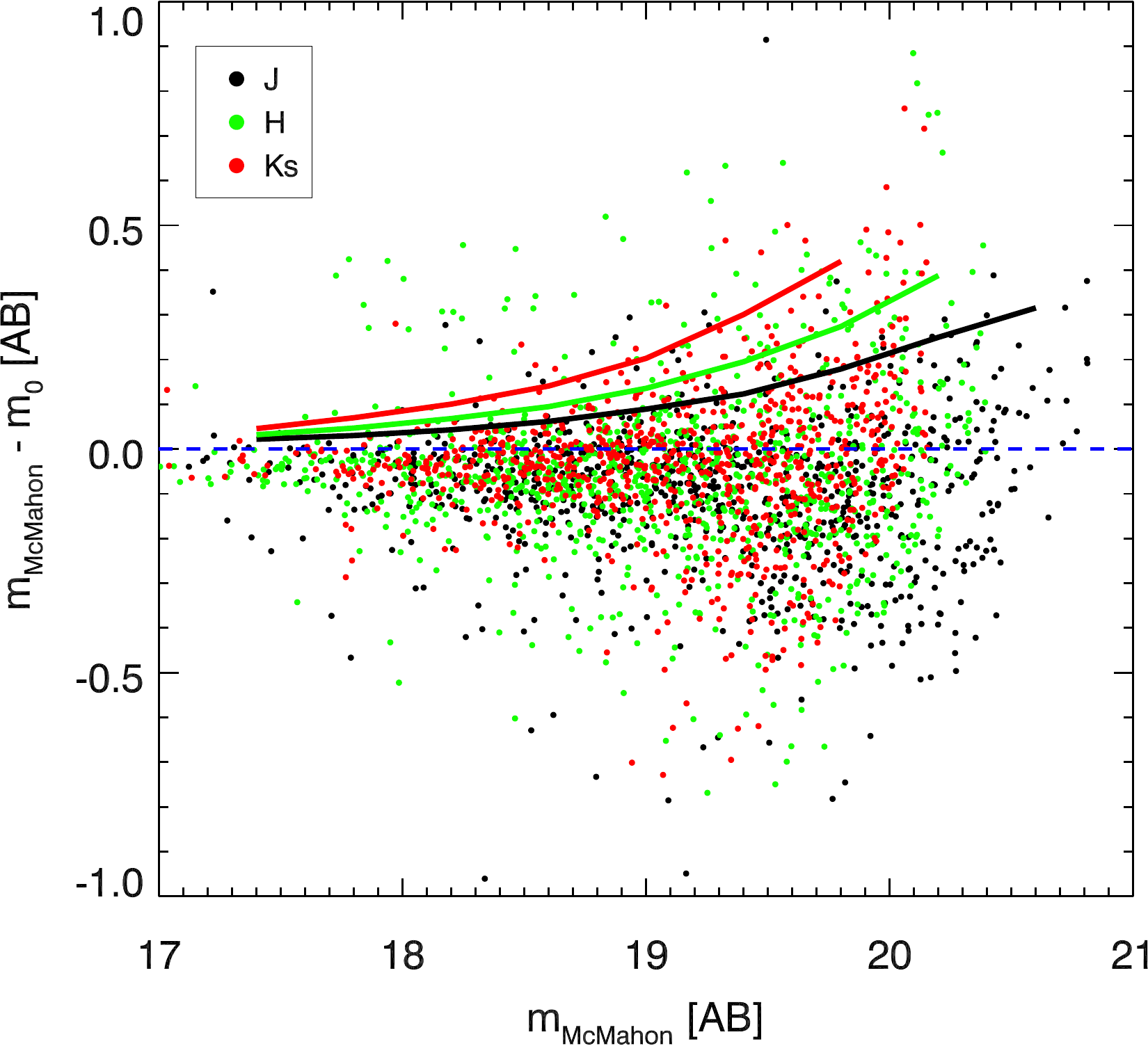}}
\caption{\small Comparison between our VHS-band measurements (m$_0$) and measurements from the catalog that came with the VHS data \citep[m$_{\text{McMahon}}$;][]{Mcmahon2012} for a random subset of $\approx$ 900 stars. $J$ is shown in \textit{black}, $H$ in \textit{green}, and $K_s$ in \textit{red}. Solid lines represent the mean y-axis errors (shown as an offset from 0), as a function of m$_{\text{McMahon}}$ in bins of 1 mag. These represent the uncertainty expected in comparing the two catalogs. \vspace{2mm}}
\label{fig:vhscompare}
\end{figure}

To confirm the reliability of our measurements, we compare our  $J$, $H$, and $K_s$ magnitudes with the source catalog released with the public VHS data. In particular, we  select stars from our catalog using Equation (\ref{eq:stars}), with $<$ instead of $\geq,$ and SExtractor \texttt{FLAGS} = 0 and match them with a random 10,000 source subset of the pre-made VHS catalog, where we define a match as  two sources within 0.5 arcsec of each other.  Note that our magnitudes are measured within 3 arcsec diameter apertures while the pre-made catalog uses $2.83$ arcsec diameter apertures.
A plot of the difference in magnitudes between our catalog and the pre-made VHS catalog can be seen in Figure \ref{fig:vhscompare}. To remove extreme outliers, magnitudes from both catalogs are cut at the depths given in Table \ref{tab:bands}. The mean offsets from 0 in the magnitude differences are -0.11, -0.05, and -0.03 mag for $J$, $H$, and $K_s$, respectively. The mean photometric uncertainty in those offsets across all magnitudes are $\pm$0.09, $\pm$0.12, and $\pm$0.18 mag, respectively. The solid lines in Figure \ref{fig:vhscompare} represent the mean uncertainty within magnitude bins of width 1 mag, and they are plotted as positive offsets from 0 on the $y$-axis. The uncertainty includes any differences in the measurement process between this paper and \citet{Mcmahon2012}, as well as the inherent uncertainty in the SExtractor measurements. As can be seen, the difference in magnitudes is reasonably within the uncertainty.

\begin{figure*}[t]
\centerline{\includegraphics[height=14.cm]{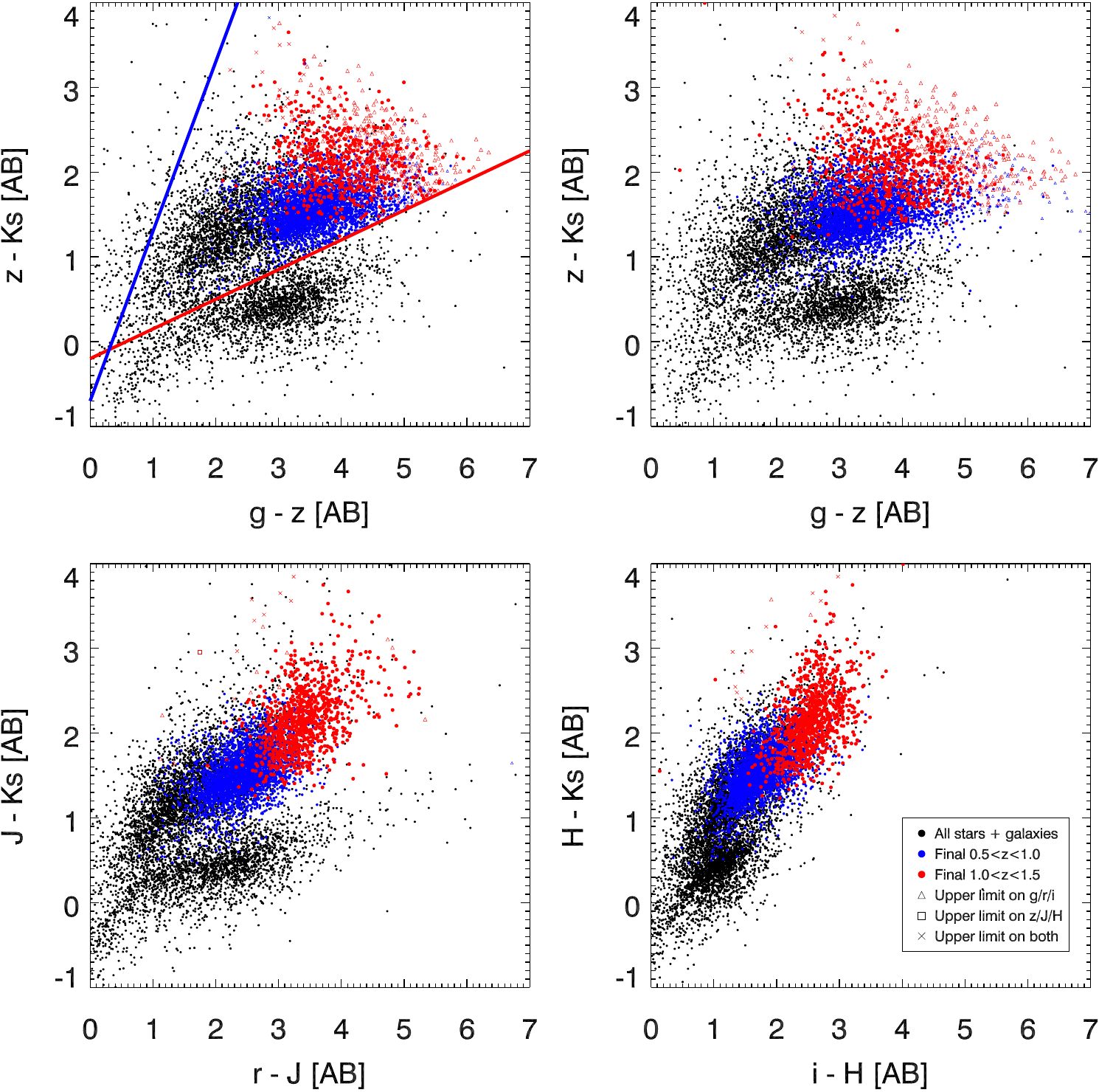}}
\caption{\small Color-color plots of a random 1/50th of the total sources described in Section \ref{sec:photfit} (\textit{black}), our final $0.5 \leq z \leq 1.0$ galaxies (\textit{blue}), and our final $1.0 \leq z \leq 1.5$ galaxies (\textit{red}). \textit{Upper left}: $gzK_s$ plot with no correction for Galactic dust extinction, showing the cuts described in Section \ref{sec:galselect}. Notice the clear distinction between stars (sources below the red line) and galaxies (sources above the red line). \textit{Upper right}: $gzK_s$ plot where galaxies have been corrected for Galactic dust extinction. \textit{Lower left}: $rJK_s$ plot where galaxies have been corrected for Galactic dust extinction. \textit{Lower right}: $iHK_s$ plot where galaxies have been corrected for Galactic dust extinction. \vspace{4mm}}
\label{fig:gzk}
\end{figure*}

We are not able to carry out a similar comparison for the BCS bands because the data do not come with reliable zeropoints, which are required to convert the measured image-level fluxes to actual fluxes. Instead, we compute the BCS band zeropoints ourselves using the stellar locus regression (SLR) code Big MACS \citep{Kelly2014}\footnote{code.google.com/p/big-macs-calibrate}.  This code calibrates the photometric zeropoints by creating a model stellar locus for every input filter and fitting them simultaneously to a selection of input stars.  To input the best possible selection of stars for this purpose, we use a combination of several criteria that are fine-tuned for each tile to balance between quality and quantity. These include selecting stars using Equation (\ref{eq:stars}), SExtractor \texttt{FLAGS} = 0, \texttt{CLASS\_STAR} $\geq$ 0.9, \texttt{A\_IMAGE/B\_IMAGE} $\geq$ 0.8, \texttt{FWHM\_IMAGE} within a certain range from the point-source limit, and selecting bright, but unsaturated fluxes. 

Our star selection results in a mean of 525 stars used per tile. We run the code using the BCS bands ($g$, $r$, $i$, $z$) plus $J$ and $H$. Since the VHS bands ($J$, $H$, $K_s$) already have accurate zeropoints, we use the code to compute the zeropoints of the other 5 bands relative to $H$. This allows us to do an independent check on the code by comparing the code's value for the $J$ zeropoint with the actual $J$ zeropoint. We find that the mean difference between the two is 0.0078 mag, which is close to the uncertainty of the code. The mean uncertainties in the derived zeropoint calibrations are 0.043, 0.037, 0.018, 0.012, and 0.0052 mag for $g$, $r$, $i$, $z$, and $J$, respectively.


\subsection{Photometric Fitting}
\label{sec:photfit}

Having obtained a calibrated catalog of sources, we then apply an initial set of cuts to remove cases that are too uncertain to be suitable for stacking.   Our goal here is not to select a statistically complete set of large, old, passive, $z \geq 0.5$ galaxies in the survey area, but rather to select a subset of such galaxies that can be cleanly identified.  To count any source as reliable, we first require that it triggers no SExtractor output flags (\texttt{FLAGS} = 0).  This choice excludes: (i) sources that have neighbors bright enough to bias the photometry; (ii) sources that were originally blended with another source;  (iii) sources with at least one saturated pixel; (iv) sources with incomplete or corrupted data; and (v) sources for which a memory overflow occurred when measuring their flux.  Furthermore, we remove all sources with a measured \texttt{FLUXERR\_APER} $\leq0$ in any band, and any source within 3 $\times$ \texttt{FWHM\_IMAGE} from the edge of a tile, since the data become unreliable near these boundaries due to dithering.

\begin{figure}[t]
\centerline{\includegraphics[height=8cm]{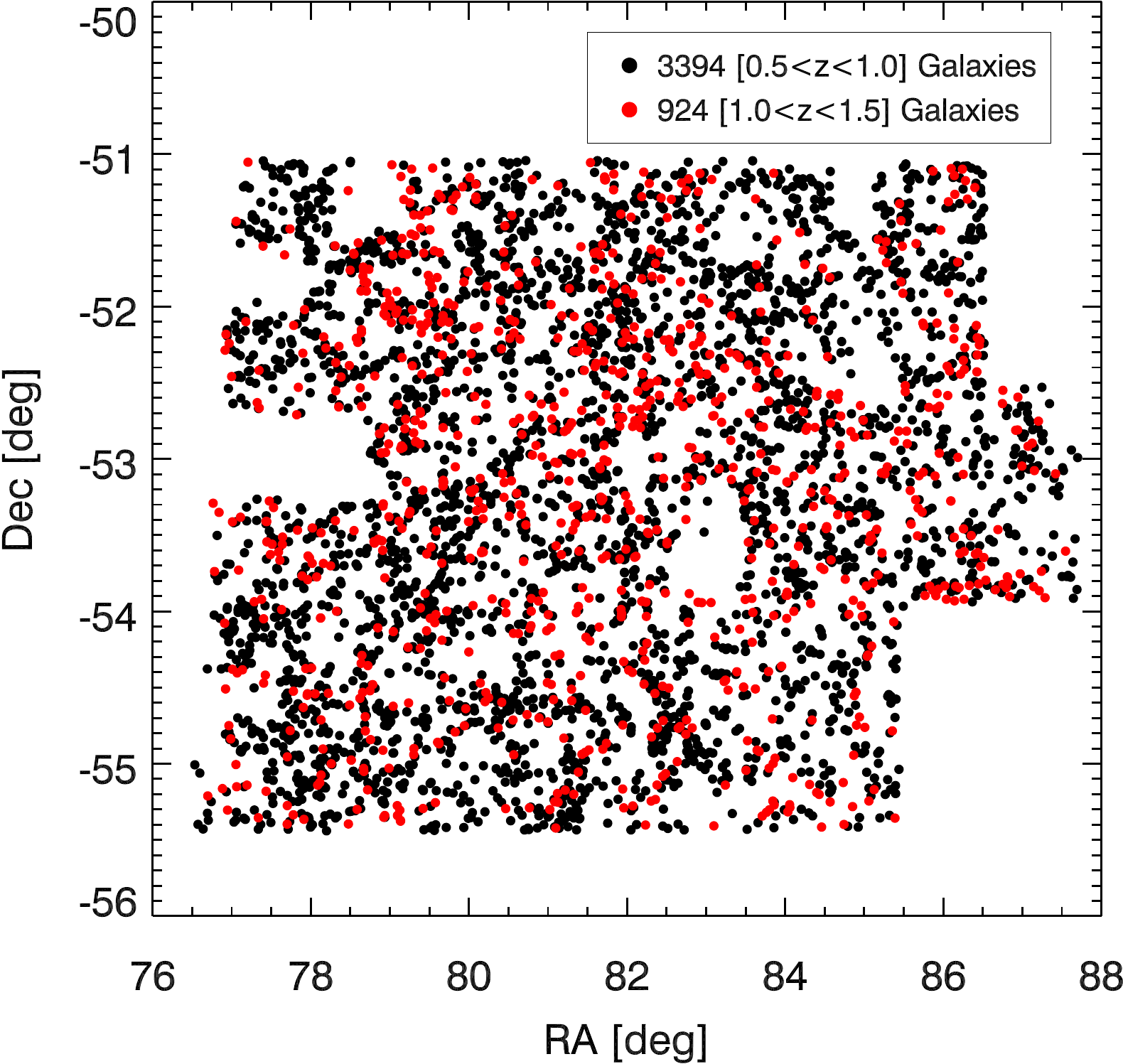}}
\caption{\small Sky distribution of our final selected galaxies for $0.5\leq z \leq 1.0$ (\textit{black}) and $1.0 < z \leq 1.5$ (\textit{red}). \vspace{3mm}}
\label{fig:skyplot}
\end{figure}

\begin{figure}[t]
\centerline{\includegraphics[height=8cm]{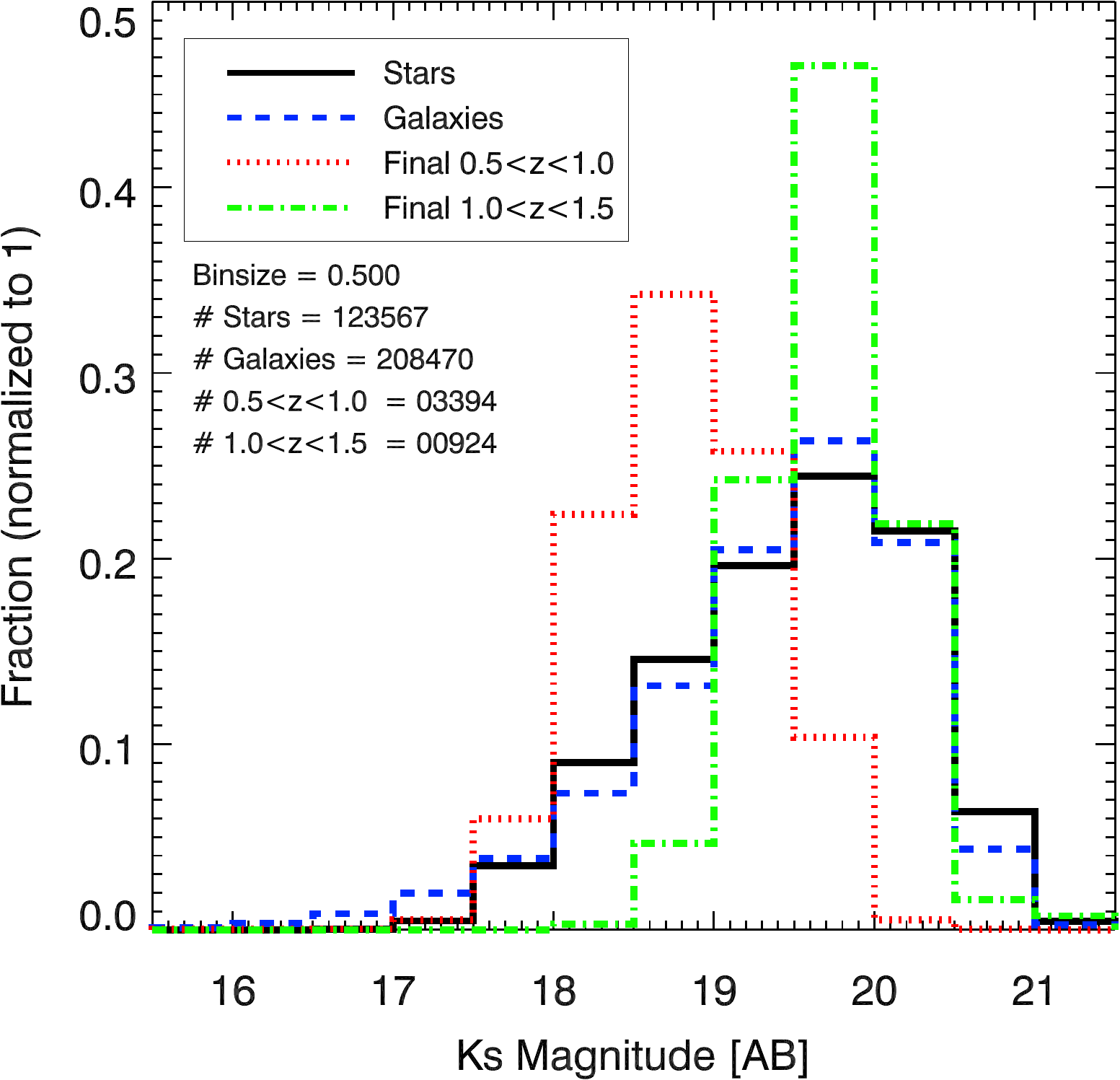}}
\caption{\small Normalized $K_s$ band magnitude histograms of our identified stars (\textit{black solid}), galaxies (\textit{blue dashed}), and final selected galaxies
at low redshift (\textit{dotted red line}) and high redshift (\textit{dotted-dashed green line}). Galaxy magnitudes have been corrected for dust extinction, as discussed in Section \ref{sec:photfit}. \vspace{3mm}}
\label{fig:khisto}
\end{figure}

Next, we separate stars from galaxies by making use of the $gzK_s$ method given by Equation (\ref{eq:stars}).   As in the plot of model galaxies (lower-right panel of Figure \ref{fig:candidates}), our data (Figure \ref{fig:gzk}) shows a clear division between the galaxy locus and the stellar locus along this limit.   Note however that  \citet{Arcila-Osejo2013} proposed a star cut of $(z-K_s) < 0.45(g-z)-0.57$, which differs from ours slightly. Furthermore, we apply Equation (\ref{eq:passivecut}) to separate out young, lower-redshift galaxies from the $z\geq 0.5$ old, passive galaxies we are interested in. After applying these criteria, we are left with a catalog of 332,037 sources consisting of 123,567 stars (37\%) and 208,470 galaxies (63\%), 195,426 (59\%) of which satisfy Equation (\ref{eq:passivecut}).
We then correct for Galactic dust extinction using the \cite{Schlegel1998} dust map and the extinction curve of \cite{Fitzpatrick1999}. Source-count histograms of the $K_s$ magnitudes for stars and the corrected $K_s$ magnitudes for galaxies are shown by the solid black and dashed blue lines, respectively, in Figure \ref{fig:khisto}.

With our catalog of galaxies we use the EAZY software package  \citep{Brammer2008} to estimate photometric redshifts and the FAST software package \citep{Kriek2009} to estimate various characteristics such as redshift, age, mass, and star formation rate (SFR). First, EAZY steps through a grid of redshifts, fits linear combinations of template spectra to our photometric data, and ultimately finds the best estimate for redshift, including optional flux- and redshift-based priors. We  allow for fits to make use of linear combinations of up to two of the default template spectra, and also apply the default $K$-band flux- and redshift-based prior derived from the  GOODS-Chandra Deep Field-South \citep{Wuyts2008}.

The resulting redshifts are then fed into the FAST code, along with our seven-band photometric data, to fit for six additional parameters: age, mass, star formation timescale $\tau$, SFR, dust content, and metallicity.  FAST allows for a range of parameters when generating  model fluxes, and in this analysis we choose: (i) a stellar population synthesis model as in \cite{Conroy2010}; (ii) a \cite{Chabrier2003} initial stellar mass function; (iii) an exponentially declining star formation history $\propto \exp(-t/\tau)$; and (iv) a dust extinction law as given by  \cite{Kriek2013}.  To determine the best-fit parameters, the code simply determines the $\chi^2$ of every point of the model cube and finds the minimum.
While the code allows for confidence intervals calibrated using Monte-Carlo simulations, here we simply make use of the best-fit values for each galaxy, recording its $\chi^2$ for use in our final galaxy selections, described in Section \ref{sec:fingal}.


\section{Final Galaxy Selection}
\label{sec:fingal}

\begin{figure*}[t]
\centerline{\includegraphics[height=6cm]{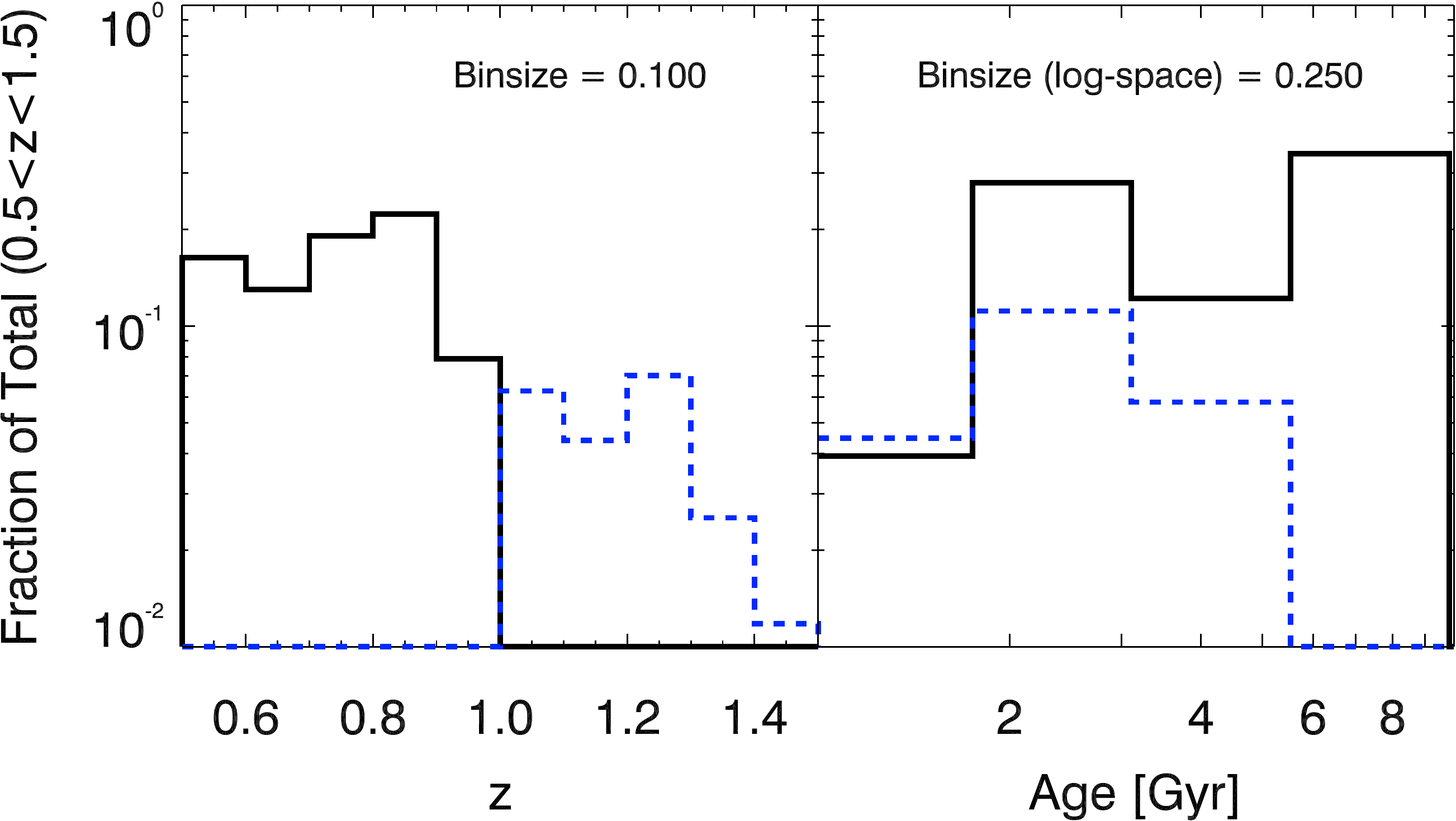} \quad
\includegraphics[height=6cm]{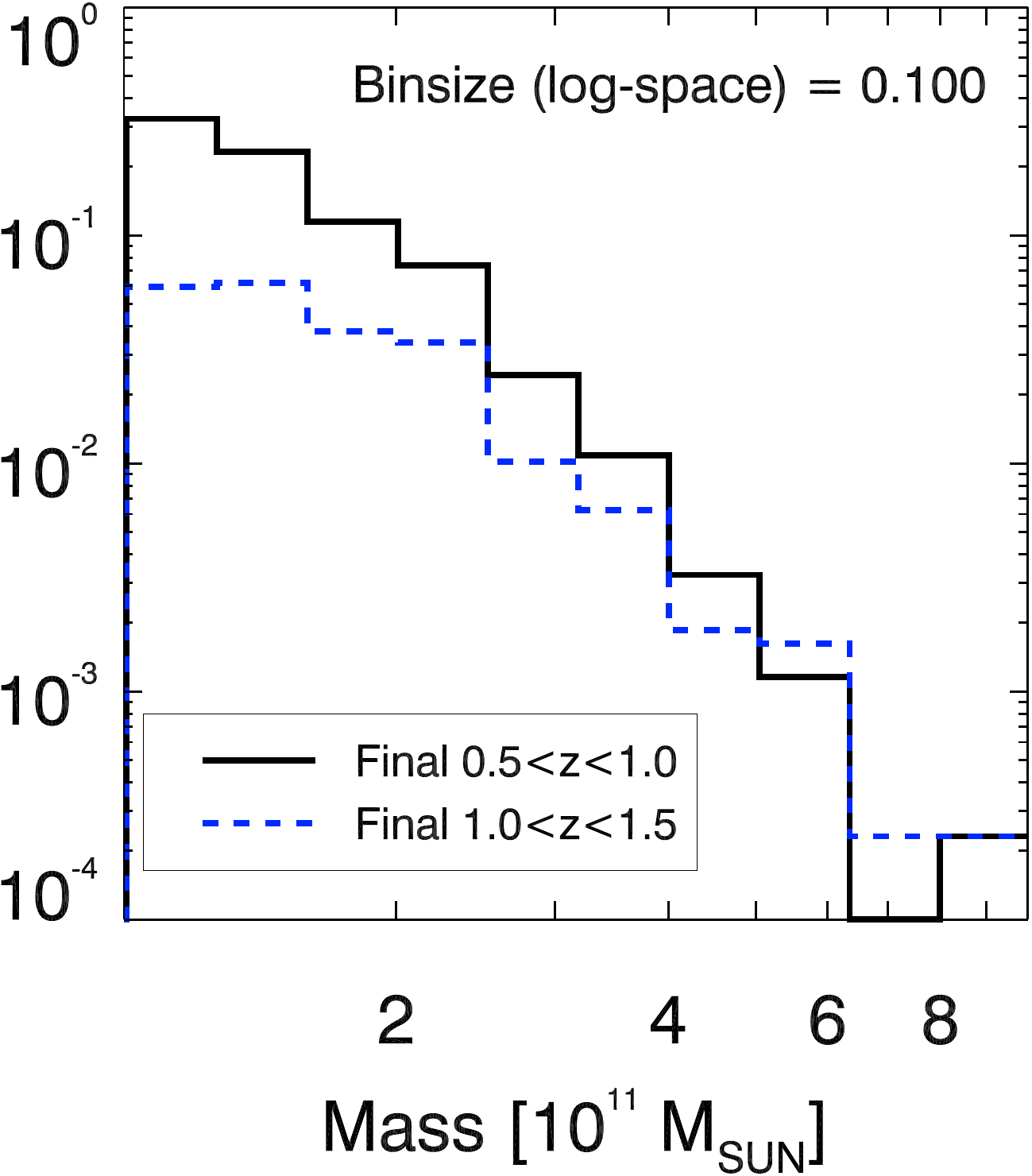}}
\caption{\small Redshift, age, and mass distributions for our final $z=0.5-1.0$ (\textit{black solid lines}), and $z=1.0-1.5$ (\textit{blue dashed lines}) galaxies. \vspace{3mm}}
\label{fig:histos}
\end{figure*}

To select the final galaxies used to measure the tSZ signal, we first cut out the least reliable FAST model fits by requiring $\chi^2 \leq 5$. Motivated by Sections \ref{sec:gas} and \ref{sec:galselect}, we then select galaxies with ages $\geq$ 1 Gyr and masses $\geq 10^{11} M_\odot$. To remove any presently star-forming galaxies, we also require the specific star formation rate SSFR $ \equiv$ SFR/mass $\leq 0.01$ Gyr$^{-1}$ \citep{Kimm2012}. This insures that we  select massive, old, and quiescent galaxies. We further split the resulting galaxies into two redshift ranges, $0.5\leq z \leq 1.0$ (``low-$z$'') and $1.0 < z \leq 1.5$ (``high-$z$''). Applying these constraints results in 4537 galaxies at low-$z$ and 1259 galaxies at high-$z$.

Our final step is to remove any galaxies known to be a likely contamination to the tSZ signal we are interested in, due to: (i) the presence of a dusty Galactic molecular cloud;
(ii) the presence of an AGN; (iii) the presence of a galaxy with strong dust emission; or (iv) the presence of a galaxy cluster, such that the tSZ signal would be dominated by the intracluster medium rather than the circumgalactic medium in which we are interested. Thus, we cut out any galaxy that is within 4 arcmin of any source found within a large number of external source catalogs, chosen to remove all such sources. Regarding these external source catalogs, to exclude the presence of Galactic molecular clouds, we remove sources correlated with the \textit{Planck} Catalogue of Galactic Cold Clumps \citep{PlanckCollaboration2015c}. To exclude the presence of bright AGN, we remove sources correlated with the \textit{ROSAT} All-Sky Survey Bright Source, Correlation, and Faint Source Catalogs \citep{Voges1999}.  To exclude strong dust emitting sources, we remove sources correlated with the \textit{Planck} Catalogue of Compact Sources \citep{PlanckCollaboration2014}, the SPT-SZ Point Source Catalog \citep{Mocanu2013},  the  AKARI/FIS All-Sky Survey Bright Source Catalogue \citep{Yamamura2010}, the AKARI/IRC All-Sky Survey Point Source Catalog \citep{Ishihara2010}, the IRAS Point Source Catalog \citep{IRAS1988},  and all sources classified as Hot DOGs  from the \textit{WISE} All-Sky Data Release Source Catalog \citep{Wright2010}. Hot DOGs are defined as sources detected in \textit{WISE} bands W3 or W4 but not in either W1 or W2 \citep[e.g.,][]{Eisenhardt2012}. Finally, to exclude sources in galaxy clusters, we remove sources correlated with the \textit{Planck} Catalogue of Sunyaev-Zeldovich Sources \citep{PlanckCollaboration2015b} and the SPT-SZ Cluster Catalog \citep{Bleem2015}.

We also carry out additional co-adds removing sources correlated with three radio surveys in addition to the cuts above, in order to further exclude potential bright AGNs. These are the Australia Telescope 20 GHz Survey Source Catalog \citep{Murphy2010}, the Parkes-MIT-NRAO (PMN) Southern Survey Source Catalog \citep{Wright1994}, and the Sydney University Molonglo Sky Survey (SUMSS) Source Catalog \citep{Mauch2003}. We find that these additional cuts do not significantly change our results, as explained further in Section \ref{sec:coadd}.

The purpose of using all of these external catalogs is to increase the reliability of our galaxy catalog, which we maximize by aggressively using every external source catalog relevant for potential contamination. This process is imperfect, though, due to the completeness limits of the external catalogs we use as well as the restriction of only using existing publicly available catalogs. We implicitly account for the residual contamination left over from our imperfect contamination removal in Section \ref{sec:sptcont}, where we model what the impact of this undetected contamination is on our measurements.

Applying these cuts results in our final sample of galaxies: 3394  at low-$z$ and 924 at high-$z$. Their distribution on the sky is shown in Figure \ref{fig:skyplot}, where we can perhaps start to see signs of large-scale structure.  Histograms of the $K_s$ magnitudes for these final two groups are shown  in Figure \ref{fig:khisto}. Their locations in color-space are plotted in Figure \ref{fig:gzk}.  Several things are evident in this figure. First, in the $gzK_{s}$-plot the stars clearly  separate out  from the galaxies  (red line and  Equation (\ref{eq:stars})), showing the quality  of our photometry. Secondly, we can see that the  blue line (Equation (\ref{eq:passivecut})) used to pare down the sample and select old and quiescent galaxies is, in  fact, a conservative cut  with respect to the  results of the SED fitting, i.e., there are very few red or blue points near the blue line. The upper two plots show the results before and after correction for Galactic extinction \citep{Schlegel1998}. These plots show that making the color-cuts before applying  this correction does not introduce any substantial contamination of our  final sample that is selected after the SED fitting stage.

Figure \ref{fig:histos} shows the redshift, age, and mass distributions of our final galaxy selection.   We can see that the number of galaxies as a function of mass is dominated by the lowest mass galaxies, although there are fewer of the fainter, lower mass galaxies detected in the higher redshift range.  Notice also that the oldest galaxies are found in the lower redshift bin, as expected.  To allow for ease of comparison between our results and theoretical models, mean values for redshift, $l_{\text{ang}}^2$, mass, age, and $K_s$-band luminosity, as well as mass-averaged values of redshift and $l_{\text{ang}}^2$, are shown for both redshift subsets in Table \ref{tab:meanvals}.   The mean and mass-averaged redshifts and luminosity distances are very close to each other, indicating no strong evolution of the mass distribution within each redshift bin.

\begin{table*}[t]
\begin{center}
\resizebox{12cm}{!}{
\begin{tabular}{|c|c|c|c|c|c|c|c|}
	\hline
	$z$ & $\left< z \right>$ & $\left< l_{\text{ang}}^2 \right>$ & $\left< \text{M} \right>$ & $\left< \text{Age} \right>$ & $\left< \text{L}_{Ks} \right>$ & $\left< z \right> _\text{M}$ & $\left< l_{\text{ang}}^2 \right> _\text{M}$ \\
	  &  & (Gpc$^2$) & (M$_\odot$) & (Gyr) & (erg s$^{-1}$ Hz$^{-1}$) &  & (Gpc$^2$) \\ \hline
	  $0.5-1.0$ & 0.72 & 2.30 & 1.51 $\times$ 10$^{11}$ & 4.34 & 2.78 $\times 10^{30}$ & 0.72 & 2.30 \\
	  $1.0-1.5$ & 1.17 & 3.02 & 1.78 $\times$ 10$^{11}$ & 2.64 & 4.07 $\times 10^{30}$ & 1.19 & 3.03 \\ \hline
\end{tabular}
}
\end{center}
\caption{\small Mean and mass-averaged values for several relevant galaxy parameters in the two final redshift ranges.}
\label{tab:meanvals}
\end{table*}


\section{Measuring the tSZ Signal}
\label{sec:tszsig}


\subsection{SPT-SZ Filtering}
\label{sec:filter}

\begin{figure}[t]
\centerline{\includegraphics[height=8cm]{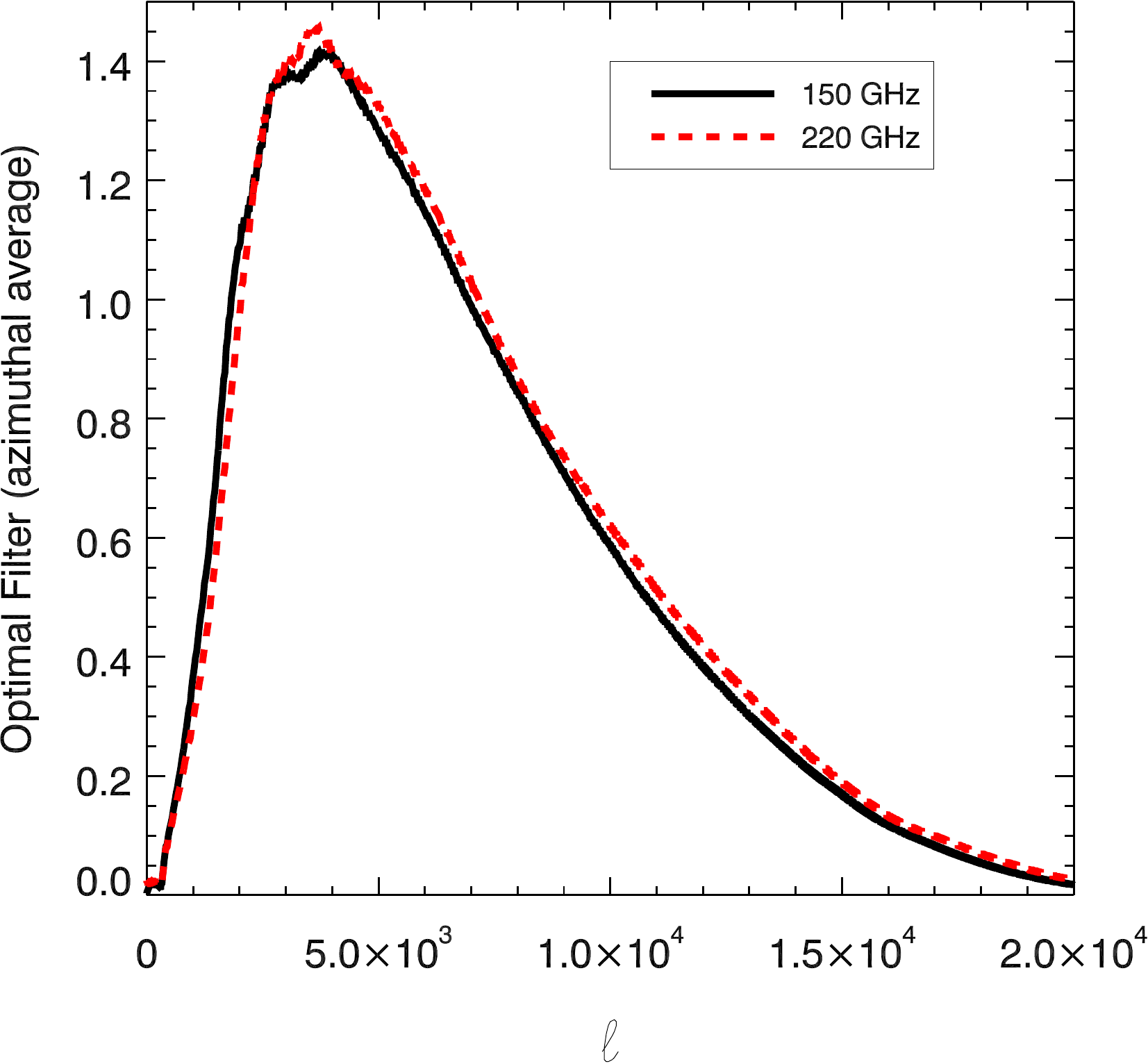}}
\caption{\small Optimal azimuthally-averaged filter curves in $\ell$-space for both SPT bands. These are scaled to preserve the flux within a 1 arcmin radius circle in the SPT images. \vspace{5mm}}
\label{fig:filt}
\end{figure}

As discussed in Section \ref{sec:gas}, the signal we are looking for occurs on arcminute scales, comparable to the resolution of the SPT-SZ data we are working with. On the other hand, the overall anisotropy of the CMB is dominated by the primary signal, which is strongest on degree scales.
For this reason it is essential for us to filter our maps before obtaining our measurements.
Since we are making measurements on the smallest scales (approaching the beam size), we apply a filter to the SPT-SZ data in order to optimize point-source measurements. This optimal filter in Fourier-space, $\psi$, is \citep{Schaffer2011}
\begin{equation}
\psi = \frac{\tau}{P} \left[ \int d^2k \frac{\tau^2}{P} \right]^{-1} ,
\label{eq:psi}
\end{equation}
where $\tau$ is the Fourier-space source profile and $P$ is the Fourier-space noise power spectrum, which is the  sum of the (squared) instrument-plus-atmosphere power spectral density and the primary CMB power spectrum.  For a point source $\tau = B \times F$, where $B$ is the Fourier-space beam function and $F$ is the Fourier-space filter transfer function. 
We then scale $\psi$ in order to preserve the total flux within a 1 arcmin radius circle in each map. Thus we expect our primary signal, which we measure in a 1 arcmin radius aperture around our stacked galaxies, to be minimally affected by our filtering. The resulting optimal point source filters for the 150 and 220 GHz bands are shown in Figure \ref{fig:filt}.

\pagebreak
\subsection{Galaxy Co-adds}
\label{sec:coadd}

We carried out our final co-add measurements by averaging the SPT-SZ maps around the galaxies in both our final low- and high-redshift galaxy samples.   Before we are able to measure a signal from these averages, however, we first need to correct for a bias introduced by our removal of all sources within 4 arcmin of contaminating sources.   Because the SPT-SZ maps themselves are normalized to a mean of 0, and all of the contaminating sources introduce positive signal into the maps, the average value in the uncontaminated regions of the maps is slightly biased to negative values. We therefore calculate a bias for the ``contaminant-free'' images by choosing 140,000 random points in our field (chosen so that there are not more random points than possible beams on the sky) and subjecting the points to the same contaminating-source cuts as our galaxies. We then take the resulting 107,561 random points and compute the mean sums within a 1 arcmin radius around each point. With these values we calculate an offset value needed to re-normalize the mean to 0. These offset values are $0.24 \pm 0.09$ and $0.58 \pm 0.13$ $\mu$K arcmin$^2$ at 150 and 220 GHz, respectively.

\begin{figure*}[ht]
\centerline{\includegraphics[height=6.5cm]{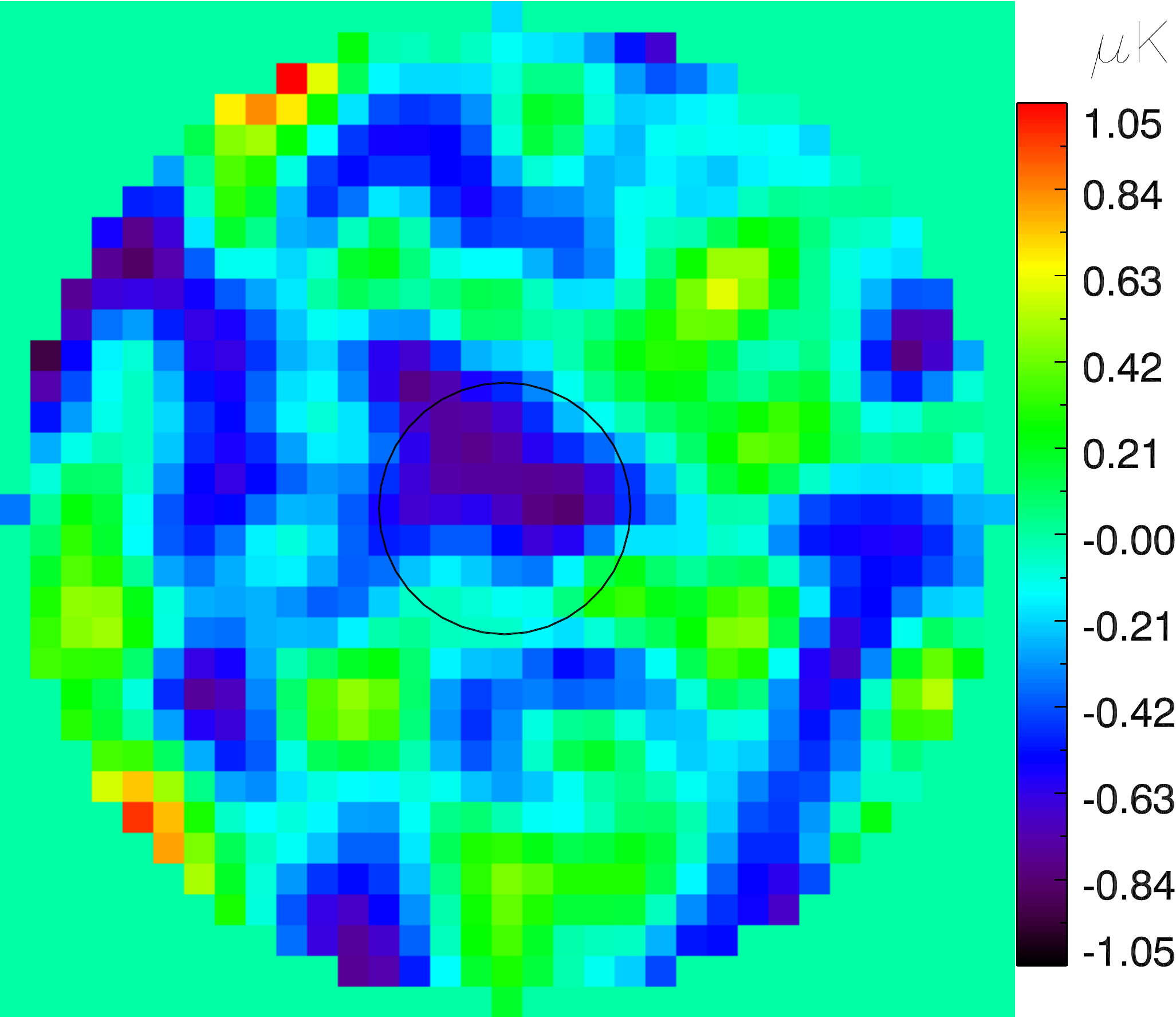} \quad
\includegraphics[height=6.5cm]{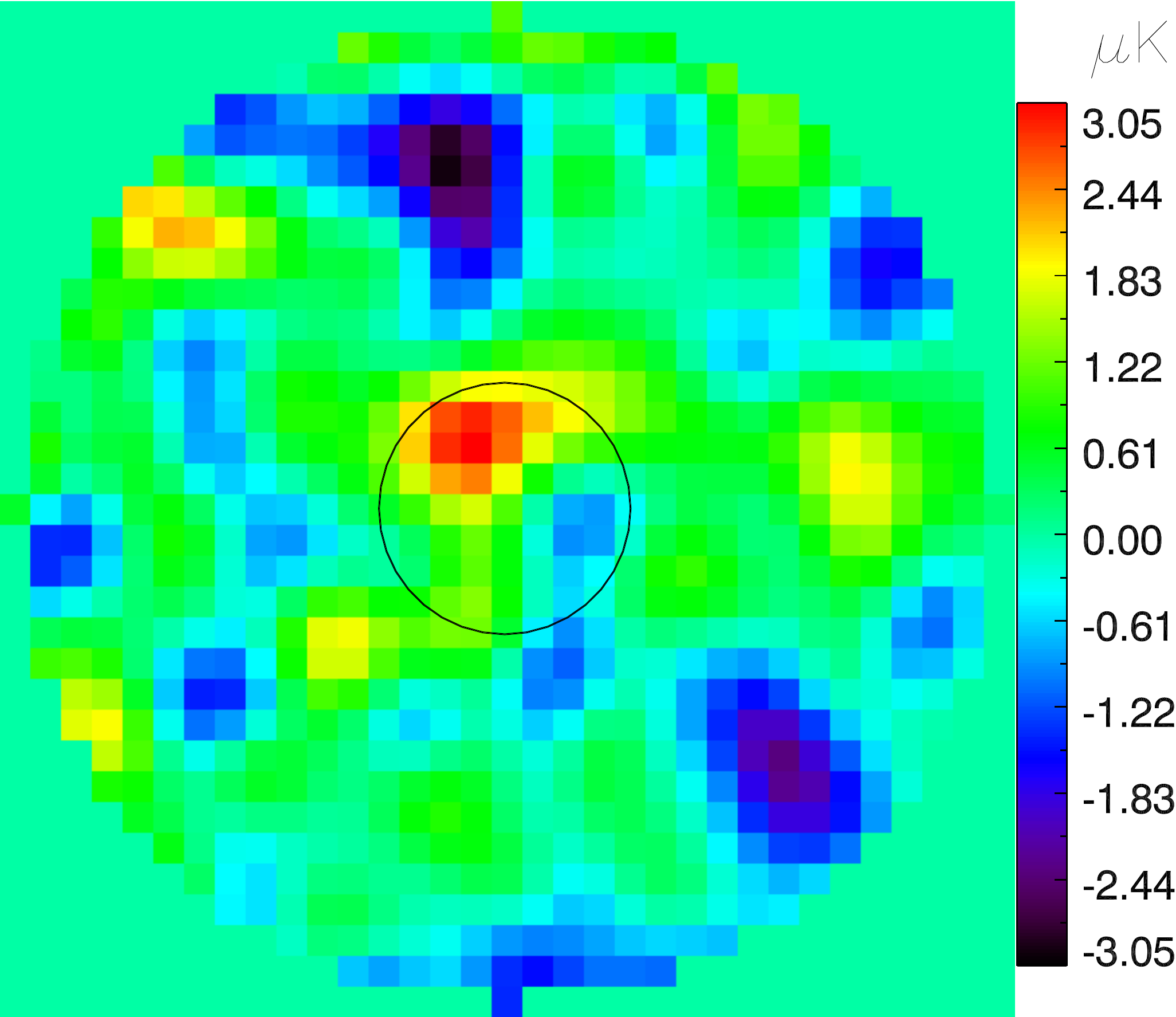}}
\centerline{\includegraphics[height=6.5cm]{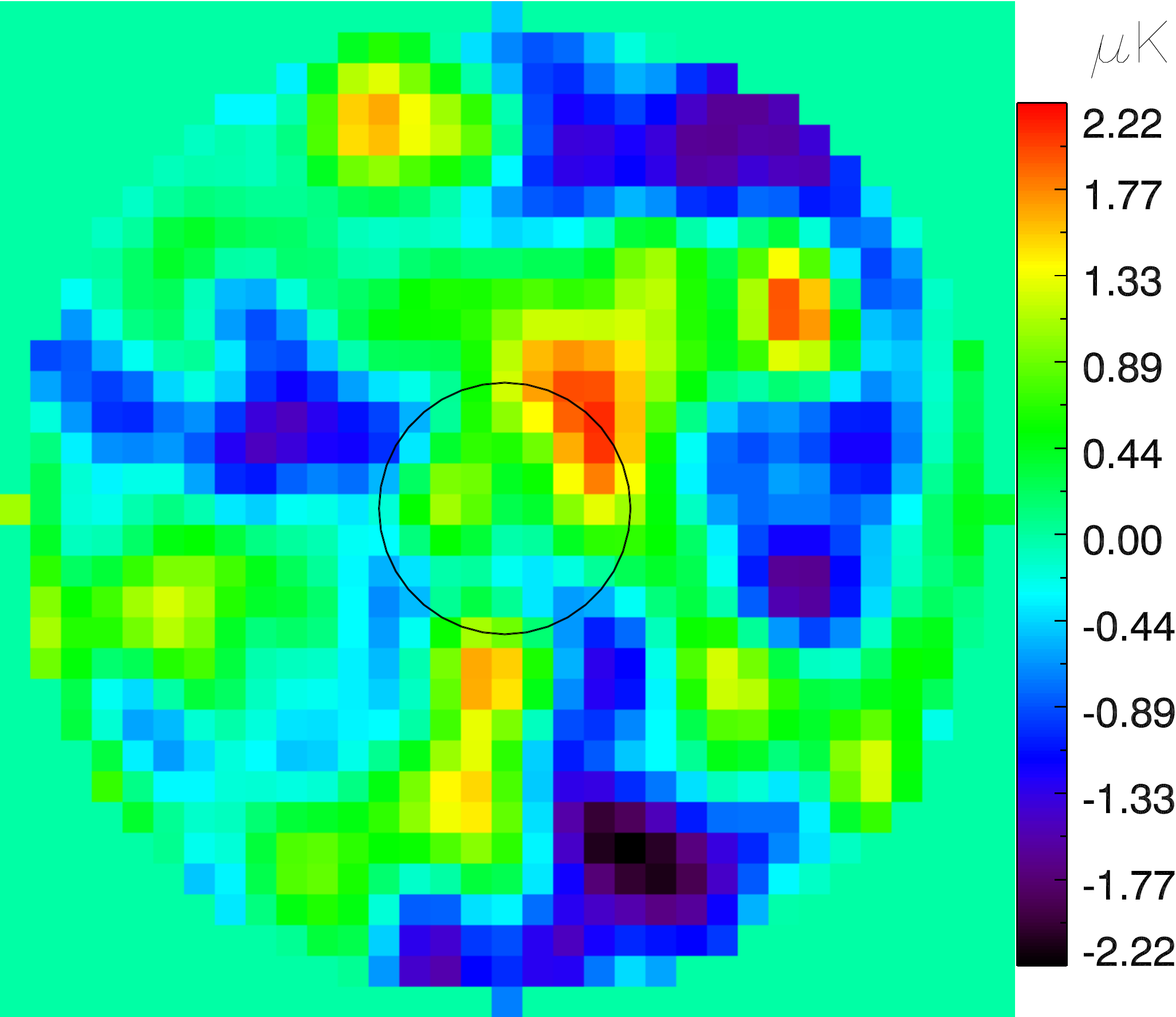} \quad
\includegraphics[height=6.5cm]{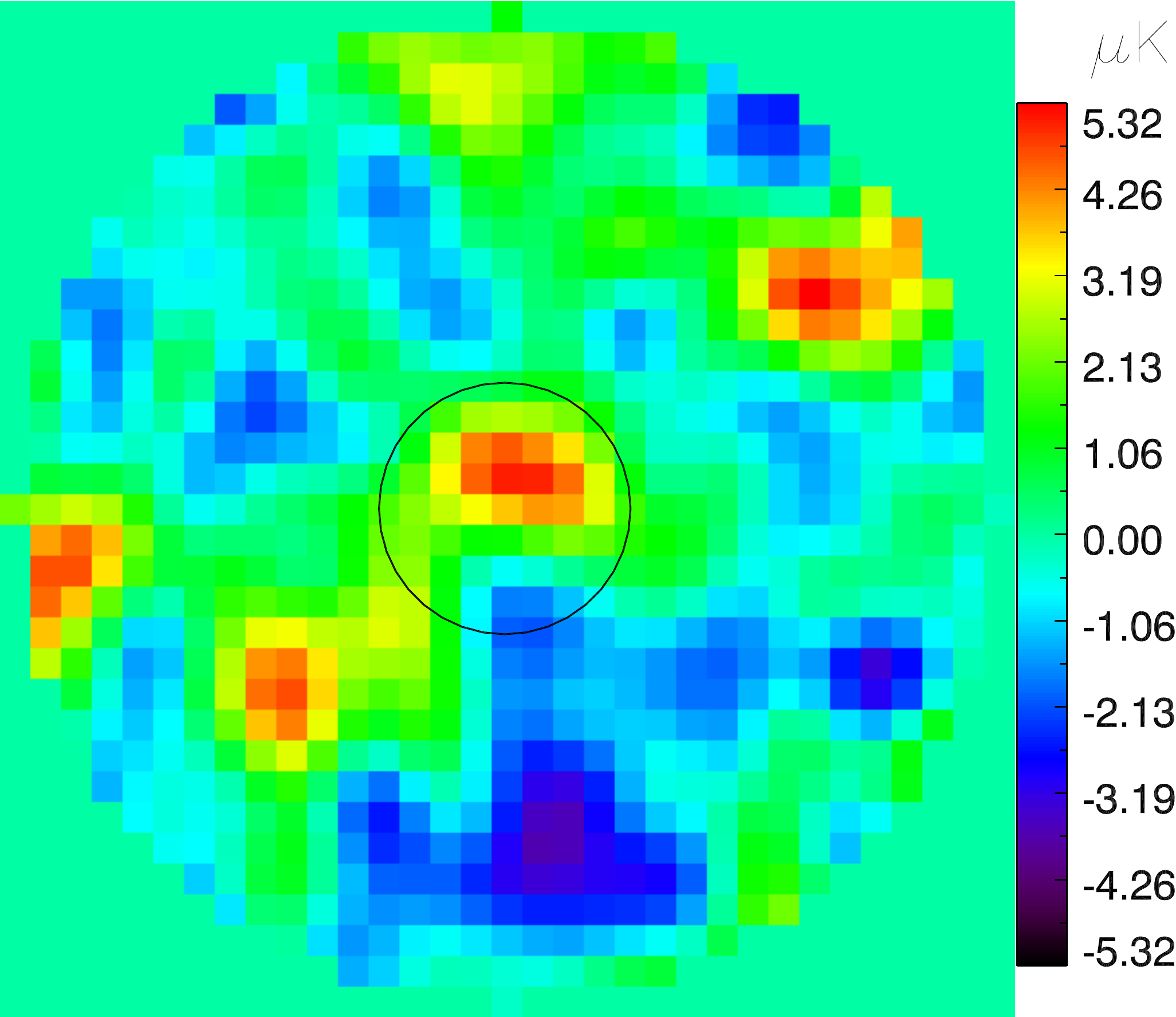}}
\caption{\small Final co-added galaxy images. \textit{Left}: 150 GHz. \textit{Right}: 220 GHz. \textit{Top}: $0.5 \leq z \leq 1.0$. \textit{Bottom}: $1.0 \leq z \leq 1.5$. The images are 8 $\times$ 8 arcmin (33 $\times$ 33 pixels). They represent the region where we have rejected any contaminating sources (see Section \ref{sec:fingal}). The black circles represent a 1 arcmin radius aperture. \vspace{3mm}}
\label{fig:coaddims}
\end{figure*}

We then sum and average the total signal within 0.5, 1, 1.5, and 2 arcmin radius apertures around our sources and add the offset, scaling them appropriately for the different aperture sizes. The 0.5 arcmin radius aperture represents roughly the size of the 150 and 220 GHz beam FWHMs, which are 1.15 arcmin and 1.05 arcmin, respectively. Additionally, we calculate the standard deviation for each of these measurements by finding the standard deviation of the same size co-added region around an equal number of random points in our field, subjected to the same contaminating source cuts. The offset uncertainties are also included but are negligible. The final co-add values for each aperture size and redshift range are given in Table \ref{tab:coadds}. The final galaxy co-add images for both redshift ranges are shown in Figure \ref{fig:coaddims}.

\begin{figure*}[t]
\centerline{\includegraphics[height=7cm]{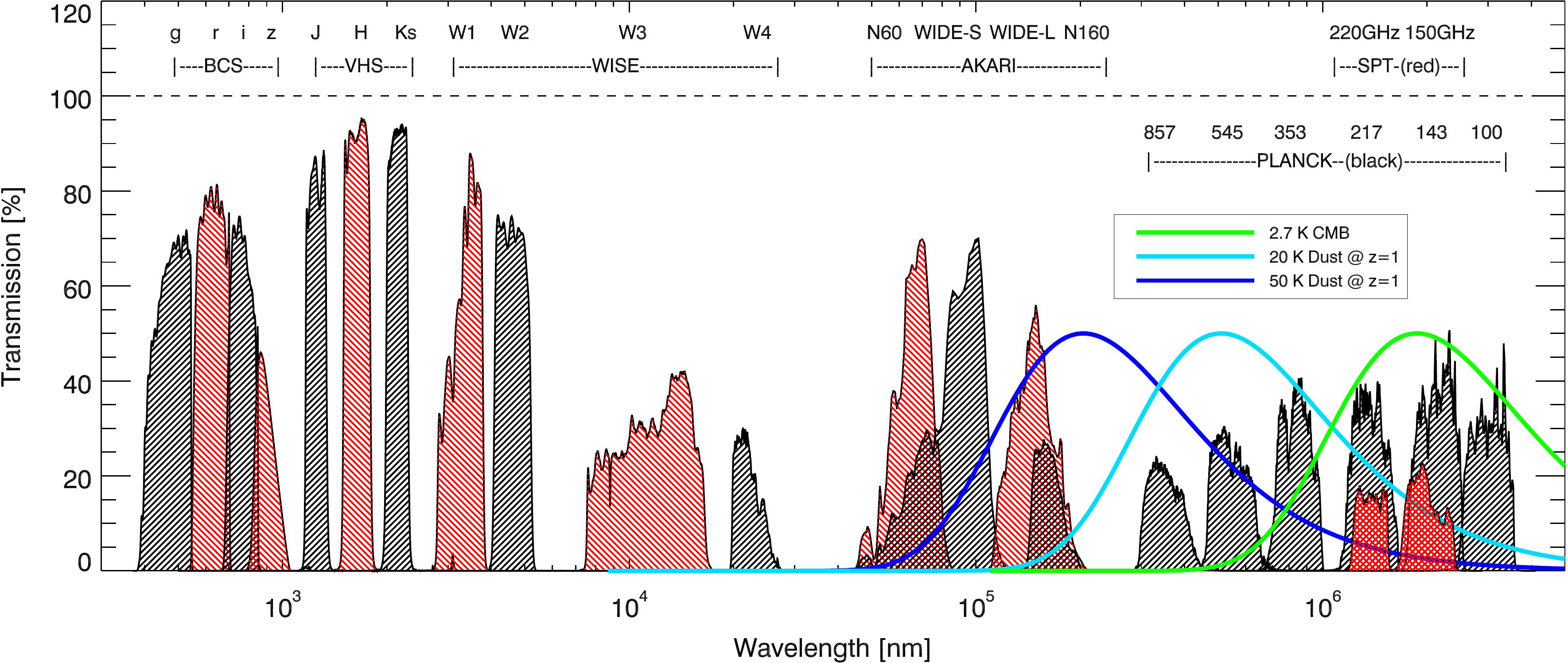}}
\caption{\small The filter curves for several of the data sets used in this paper are shown. From left to right: BCS and VHS bands used for galaxy selection, \textit{Wide-Field Infrared Survey Explorer} (\textit{WISE}), \textit{AKARI}, and \textit{Planck} bands used for identifying and constraining the signal from dusty contaminating sources, and SPT-SZ bands used for measuring the tSZ effect. The first four surveys alternate between \textit{black} and \textit{red} for each band, while \textit{Planck} bands are all \textit{black} and SPT-SZ bands are all \textit{red} to distinguish between the two. Also shown are blackbody curves for the CMB (\textit{green}), 20 K dust at $z = 1$ (\textit{light blue}), and 50 K dust at $z = 1$ (\textit{dark blue}), all normalized to 50\% on the plot. The horizontal dashed line indicates 100\% transmission. \vspace{3mm}}
\label{fig:filters}
\end{figure*}

The upper left panel of this figure shows a clear $\approx 1$ arcmin size $\approx 2\sigma$ negative feature centered directly on our stack of low-$z$ galaxies, with a magnitude consistent with a significant  tSZ signal.  Moreover, the low-$z$ 220 GHz measurements show a strong positive signal centered on our co-added sources.  
Because the tSZ effect has a negligible impact at this frequency, this indicates that  despite our cuts on detected contaminating sources, there still remains a significant positive contaminating signal at 220 GHz, made up of the sum of fainter sources. Looking at the emission by the typical range of dust temperatures at $z=1$ (light blue and dark blue curves in Figure \ref{fig:filters}), the CMB spectrum (green curve in Figure \ref{fig:filters}), and the frequencies of our SPT measurements (rightmost red hatched regions in Figure \ref{fig:filters}), it appears likely that this contaminating signal at 220 GHz extends into the 150 GHz band, meaning that the true tSZ signal for the lower-redshift galaxies we have selected is even more negative than the values in Table \ref{tab:coadds}.

Moving to the higher redshift stacks shown in the bottom panel of Figure \ref{fig:coaddims}, we find that the stacked emission of our galaxies in the 150 GHz map is consistent with zero signal.  However, as this band measures the sum of the (negative) tSZ and the (positive) contaminating sources, it is difficult to interpret these results without also considering the high-$z$ measurements at 220 GHz.   As in the lower redshift case, this band shows a clear excess, but now at a magnitude that is roughly twice that seen 
in the low-$z$ stack.  This suggests that, because it is more difficult to identify contaminating sources at higher redshift, the high-$z$ 150 GHz measurement is more contaminated than the lower-redshift measurement, covering up the negative signal in which we are interested.    In both redshift ranges, however, it is clear that obtaining the best possible constraints on AGN feedback requires making the best possible separation between the tSZ signal and the contaminating signal, a topic we address in detail below. Finally, we can convert our co-added $\Delta T$ signal into gas thermal energy using Equation (\ref{eq:newEthrm}). These values (using a 1 arcmin radius aperture) are shown in Table \ref{tab:finvals}, under ``Data only''.

\begin{table*}[t]
\begin{center}
\resizebox{12cm}{!}{
\begin{tabular}{|c|c|c|c|c|c|}
\hline
$z$ & Band & 0.5 arcmin & 1 arcmin & 1.5 arcmin & 2 arcmin\\
  & (GHz)  &  ($\mu$K arcmin$^2$) & ($\mu$K arcmin$^2$) & ($\mu$K arcmin$^2$) & ($\mu$K arcmin$^2$)\\ \hline
 0.5 - 1.0 & 150  & -0.53 $\pm$ 0.26 & -1.5 $\pm$ 0.7 & -2.3  $\pm$ 1.3 & -2.7  $\pm$ 1.9 \\              
 0.5 - 1.0 & 220  & 0.85  $\pm$ 0.53 & 3.0  $\pm$ 1.4  & 5.4   $\pm$ 2.3 & 6.4   $\pm$ 3.0 \\     
 1.0 - 1.5  & 150 & 0.39  $\pm$ 0.49 & 1.6  $\pm$ 1.4  & 2.8   $\pm$ 2.5 & 3.2   $\pm$ 3.7 \\    
 1.0 - 1.5 & 220  & 2.6    $\pm$ 1.0 & 6.3    $\pm$ 2.7  & 7.9   $\pm$ 4.4 & 7.4   $\pm$ 5.7  \\ \hline
\end{tabular} }
\end{center}
\caption{\small Final co-added signals. The columns show four different aperture sizes by radius. The smallest aperture represents roughly the beam FWHM.}
\label{tab:coadds}
\end{table*}

As mentioned in Section \ref{sec:fingal}, we performed the same co-adding method while also removing galaxies near bright radio sources. After these additional cuts, the number of final galaxies becomes 2,219 for our low-$z$ subset and 614 for our high-$z$ subset. The resulting co-add values for a 1 arcmin radius aperture are: $-1.7 \pm 0.9$ $\mu$K arcmin$^2$ for 150 GHz low-$z$, $2.4 \pm 1.7$ $\mu$K arcmin$^2$ for 220 GHz low-$z$, $0.9 \pm 1.7$ $\mu$K arcmin$^2$ for 150 GHz high-$z$, and $10.4 \pm 3.3$ $\mu$K arcmin$^2$ for 220 GHz high-$z$. We find that the additional radio source cuts do not significantly change our results except for an increased positive signal at 220 GHz in the high-$z$ subset, though they do increase the noise in our measurements because we are co-adding a smaller number of galaxies. As a result, we do not use the radio source cuts in our modeling and analysis below, though the higher 220 GHz signal in the high-$z$ subset may suggest a more significant tSZ detection in our high-$z$ results.


\subsection{Removing Residual Contamination}
\label{sec:sptcont}

In order to constrain the impact that undetected contaminating sources have on our tSZ measurements, we built a detailed model of contaminants
based on an extrapolation of the SPT source counts measured in \citet{Mocanu2013}.
Our approach is to extend these counts to fainter values by modeling a random population of undetected sources that follow the trend of the detected sources into the unresolved region,
which we then  relate to the contaminating signal in both our 150 and 220 GHz measurements. 

Following \citet{Mocanu2013} we separate contaminants into  synchrotron sources, which emit most at lower frequencies, and 
dusty sources, which emit most at higher frequencies.   For each source population we model the number counts as a power law, 
\begin{equation}
\frac{dN}{dS} = \frac{N_0}{S_{\rm max}} \left(\frac{S}{S_{\rm max}}\right)^{\alpha},
\label{eq:dnds}
\end{equation}
where $dN/dS$ is the number of sources between flux $S$ and $S + dS$, $N_0$ is an overall amplitude, $\alpha$ is the power-law slope,
and $S_{\rm max}$ is the flux at which all brighter sources have a 100\% completeness level in the source count catalog.
 We then compute a range of allowed source count slopes from the \citet{Mocanu2013} data,  by carrying out a $\chi^2$ fit in log-space.
 Our best-fit slopes at 150 GHz were $\alpha_s = -2.05 \pm 0.04$ for the synchrotron sources  and $\alpha_d = -2.70 \pm 0.19$ for the dusty sources.

Note that our calculated values for $\alpha_d$ are much steeper than $\alpha_s$, meaning that while the number density of detected sources is dominated by synchrotron sources, the number density of undetected sources is likely to be dominated by dusty emitters.   Note also that $\alpha_d$  and $\alpha_s$  are sufficiently steep that the number of sources diverges as $S$ goes to 0, meaning that the source count distribution must fall off below some as-yet undetected flux.   For simplicity, we model this fall-off as a minimum flux $S_{\rm min}$ below which there are no contaminating sources associated with the galaxies we are stacking.  

For any choice of $\alpha_d,$ $\alpha_s,$ and $S_{\rm min}$ (which we will call a ``source-count model''), we are then able to construct a model population of contaminating source fluxes through a four-step procedure as follows:
(i) for each model source, we randomly decide whether it is a synchrotron source or a dusty source, such that the overall fraction of detectable dusty sources to synchrotron sources matches the observed source counts;
(ii)  we then assign the source a random 150 GHz flux, $S_{\rm 150, rand},$ by inverting 
\begin{equation}
\int^{S_{\rm 150, rand}}_{S_{\rm150,  min}} dS \, \frac{dN}{dS} =  R  \int^{S_{\rm 150, max}}_{S_{\rm 150, min}} dS \, \frac{dN}{dS}, 
\end{equation}
where $R \in \left[0,1\right]$ is a random number, 
such that their overall population matches the source count slopes. This gives
\begin{equation}
S_{\rm 150, rand} = \left[ \left( 1-R \right) S_{\rm 150, min}^{\alpha+1} + R \, S_{\rm 150, max}^{\alpha+1}\right]^{\frac{1}{\alpha+1}};
\label{eq:srand}
\end{equation}
(iii) to obtain a corresponding flux for the source at 220 GHz we use the $\alpha^{150}_{220}$ spectral index distributions from \citet{Mocanu2013}, which we assume to have normalized Gaussian shapes with the properties (center, $\sigma$) = (-0.55, 0.55) for synchrotron sources and (3.2, 0.89) for dusty sources. We then randomly choose $\alpha^{150}_{220}$ values that fit these distributions and calculate the 220 GHz flux \citep[following][]{Mocanu2013} as
\begin{equation}
S_{220, \text{rand}}=S_{150, \text{rand}} \times 0.82 \times 1.43^{\alpha^{150}_{220}},
\label{eq:s220}
\end{equation}
where we use units of $\mu$K arcmin$^2$ for all $S$; and (iv) finally, if the source had a detectable 150 or 220 GHz flux, we randomly discard it with a probability chosen to match the completeness percentages in the source count catalog. 

For any single source-count model, we repeat the process 100,000 times, resulting in a large catalog of contaminating fluxes in both bands. From these, we compute the mean flux per contaminating source in each band, $\left<S_{150,\text{cont}}\right>$ and $\left<S_{220,\text{cont}}\right>$, which represents the contamination we are measuring in our stacks. To account for variations in the input parameters, we compute model contamination signals for a wide range of source-count models, with $S_{150,\text{max}} = 260 \, \mu$K arcmin$^2$. We vary both $\alpha_d$ and $\alpha_s$ from $-2\sigma$ to $+2\sigma$ in steps of $\sigma$, and we let $\log_{10}$($S_{\rm min}$) vary from $\log_{10}$(0.01 $\mu$K arcmin$^2$) to $\log_{10}$(30 $\mu$K arcmin$^2$) in steps of 0.2 in log-space.

\begin{figure*}[t]
\centerline{\includegraphics[height=10cm]{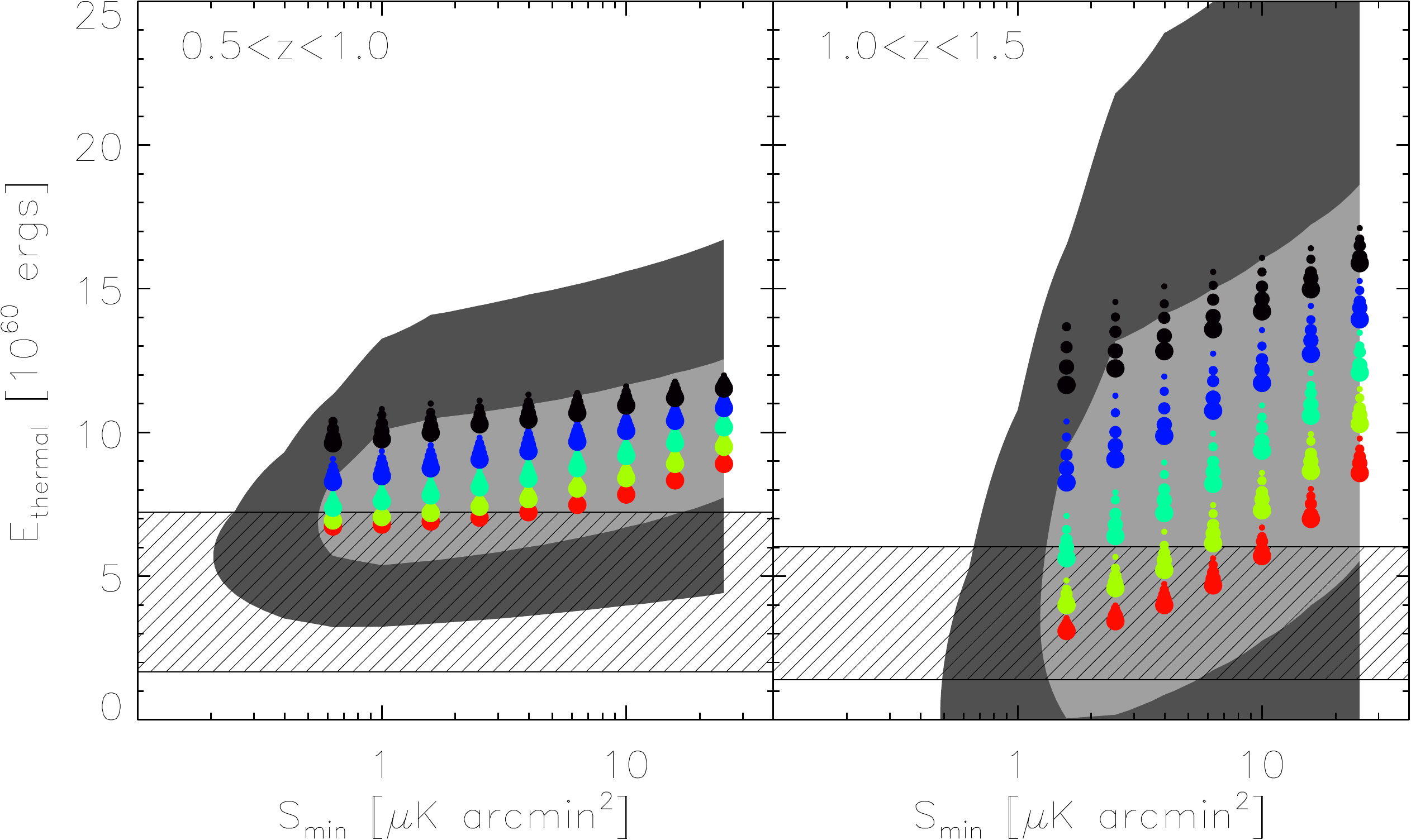}}
\caption{\small Plot of the contaminant-corrected $E_\text{therm}$ (see Equation (\ref{eq:newEthrm})) for different choices of $\alpha_{\rm dust}$, $\alpha_{\rm sync}$, and $S_{\rm min}$. Points are located at the peak $\chi^2$ probability for each model. Increasing size represents increasing (i.e. more positive) $\alpha_s$, and changing color from red to black represents increasing $\alpha_d$. The light and dark gray regions represent the complete span of $\pm1\sigma$ and $\pm2\sigma$, respectively, for all points. The hatched regions represent the $\pm1\sigma$ range for $E_\text{grav}$ [see Equation (\ref{eq:Egrav})]. \vspace{3mm}}
\label{fig:s150corr}
\end{figure*}

For each source-count model, we compute best-fit tSZ values by varying our two free parameters, tSZ signal ($S_{\text{SZ}}$) and the fraction of our measured sources that are contaminated ($f_{\text{cont}}$).   
We vary the tSZ signal from -50 to 50 $\mu$K arcmin$^2$ in steps of 0.1 $\mu$K arcmin$^2$, and we vary the fraction contaminated from -3 to 9 in steps of 0.01.
For every combination of these parameters we compute
\begin{equation}
\chi^2(f_{\text{cont}},S_{\text{SZ}}) = \mathcal{B} \times \mathcal{A}^{-1} \times \mathcal{B}^T ,
\label{eq:chi2matrixes}
\end{equation}
where $\mathcal{B}$ is the signal array,
\begin{equation}
\mathcal{B} = \left( \begin{array}{c}
f_{\text{cont}} \times \left<S_{150,\text{cont}}\right> + S_{\text{SZ}} - S_{150} \\
f_{\text{cont}} \times \left<S_{220,\text{cont}}\right> - S_{220}
\end{array} \right) ,
\label{eq:sigarray}
\end{equation}
and $\mathcal{A}$ is the noise matrix containing the noise for each band plus the covariance terms between each band,
\begin{equation}
\mathcal{A} = \left( \begin{array}{cc}
\sigma_{150}^2 & \sigma_{150}\sigma_{220} \\
\sigma_{150}\sigma_{220} & \sigma_{220}^2
\end{array} \right) .
\label{eq:noisetrix}
\end{equation}
Here, $S_{150}$, $S_{220}$, $\sigma_{150}$, and $\sigma_{220}$ are our measured 1 arcmin radius values from Table \ref{tab:coadds}. As discussed in Section \ref{sec:coadd}, the $\sigma$ values are computed using random point measurements. To be explicit here,
\begin{equation}
\begin{split}
& \sigma_i \sigma_j = \\ & \frac{\sum_{a=0}^{N_{\text{rand}}} (S^a_{i,\text{rand}}-\left<S_{i,\text{rand}}\right>)\times(S^a_{j,\text{rand}}-\left<S_{j,\text{rand}}\right>)}{N_{\text{rand}} N_{\text{source}}} ,
\label{eq:sigmaspt}
\end{split}
\end{equation}
where $i$ and $j$ represent the bands, $S_{a,\text{rand}}$ and $S_{b,\text{rand}}$ represent the 1 arcmin radius aperture values for the random points, $N_{\text{rand}} = 107,561$ is the number of random points used, and $N_{\text{source}}$ is the number of galaxies used (3394 for low-$z$ and 924 for high-$z$). We then convert the $\chi^2$ values to Gaussian probabilities $P$ by taking
\begin{equation}
\begin{split}
& P(S_{\text{SZ}}) = \\ & \frac{\sum_{f_{\text{cont}} \in [0,1]}  \exp[-\chi^2(f_{\text{cont}},S_{\text{SZ}})/2]}
		                     {\sum_{f_{\text{cont}}} \sum_{S_{\text{SZ}}}  \exp[-\chi^2(f_{\text{cont}},S_{\text{SZ}})/2]},
\label{eq:probmod}
\end{split}
\end{equation}
where the lower sum over $f_{\text{cont}}$ runs from $-3$ to $9$ and the lower sum over $S_{\text{SZ}}$ runs from $-50$ to $50$ $\mu$K arcmin$^2$.
Our approach is thus to marginalize over values of $f_{\rm cont}$ in the full physical range from $0$ to $1,$ but normalize the overall probability by the sum of $f_{\text{cont}}$ over a much larger range, including unphysical values.  This excludes models in which a good fit to the data can only be achieved by moving $f_{\text{cont}}$ outside the range of physically possible values.

Equation (\ref{eq:probmod}) then gives us a function $P(S_{\text{SZ}})$ for each combination of $\alpha_d,$ $\alpha_s,$ and $S_{\rm min}$. We can convert the corresponding $S_{\text{SZ}}$ value to the gas thermal energy, $E_\text{therm},$ using Equation (\ref{eq:newEthrm}) and the average $l^2_\text{ang}$ from Table \ref{tab:meanvals}. Note that a positive detection of the tSZ effect is seen as a negative $\Delta T$ signal at 150 GHz, and it represents a positive injection of thermal energy into the gas around the galaxy. Additionally, we compute a corresponding range for $E_\text{grav}$ using Equation (\ref{eq:Egrav}) and values from Table \ref{tab:meanvals}. The peak of each $P(S_{\text{SZ}})$ curve is shown as the colored points in Figure \ref{fig:s150corr}, where $\alpha_s$ (represented by point size) is increasing (i.e. becoming more positive) downwards, and $\alpha_d$ (represented by point color) is increasing  upwards. The 1$\sigma$ and 2$\sigma$ contours are computed for each $S_{\rm min}$ by averaging $P(S_{\text{SZ}})$ across $\alpha_d$ and $\alpha_s.$  The resulting probability distribution depends only on $E_\text{therm}$ and $S_{\rm min}$, and 1$\sigma$ and 2$\sigma$ are represented by the values $P(S_{\text{SZ}})=0.61$ and 0.13, respectively (i.e. $\exp[-\sigma^2/2]$). These contours are shown in Figure \ref{fig:s150corr}, along with the $\pm1\sigma$ range for $E_\text{grav}$. From this figure we see that there is a $>2\sigma$ tSZ detection for every source model at low-$z$, and a $\approx 1\sigma$ detection of a signal exceeding the range that can be explained without feedback. At high-$z$, where the contaminants are harder to constrain, there is a $\approx 1\sigma$ tSZ detection for every source model.

Finally, we average the probability distribution across $S_{\rm min}$ to get a final distribution as a function of only $E_\text{therm}$. The significance values of this curve are shown in the ``SPT only'' part of Table \ref{tab:finvals}. We see a $> 3\sigma$ total tSZ detection at low-$z$, and a nearly $1\sigma$ tSZ detection at high-$z$.  Furthermore, in both redshift bins, the best fit values are higher than expected from models that do not include AGN energy input.


\subsection{Removing Residual Contamination Using Planck Data}
\label{sec:planckcont}

In order to better constrain the impact of dusty contaminating sources on our measurements, we made use of the 2015 public data release from the \textit{Planck} mission, focusing on the high-frequency bands at 857, 545, 353, and 217 GHz (see the rightmost black hatched regions in Figure \ref{fig:filters}). While the FWHM beam size for this data is about 5 arcmin \citep{PlanckCollaboration2015e}, and thus  too low-resolution to detect the tSZ signal in which we are interested, the data provides information at the higher frequencies at which the dusty sources should be much brighter (i.e. the light blue and dark blue curves in Figure \ref{fig:filters}). Therefore these measurements have the potential to discriminate between contaminant models, allowing us to better remove this contribution from the tSZ signal.

Our goal is to use this data to add terms to our $\chi^2$ fit that quantify, for each source model, how consistent a given choice of $f_{\rm cont}$ is with the \textit{Planck} measurements.   To compute these extra $\chi^2$ terms, we again construct stacks of the data over each of our galaxies, but now, because of the lower resolution of the \textit{Planck} data, we  extend our contaminant source cuts to within 10 arcmin of known potentially contaminating sources. This results in a decreased number of final galaxies, now 937 at low-$z$ and 240 at high-$z$. In order to filter out the primary CMB signal, we convolve each map with a 7 arcmin FWHM Gaussian and subtract the resulting map from the original. We then stack the central pixels of every source to get co-added values for our galaxies in each of the \textit{Planck} bands. In addition, we degrade the SPT 150 and 220 GHz maps to match the beam size of \textit{Planck}, apply the same 7 arcmin FWHM filtering, and stack the galaxies on those images as well.

As was the case in Section \ref{sec:coadd},
in all of these stacks there is an offset we need to correct for since we are purposely avoiding positive contaminations in the maps. To do this we also make measurements at 3,000 random points on the sky that were restricted to the same contaminating-source cuts as our galaxies. These measurements allow us to compute offset values needed to re-normalize each band to a mean of 0, which we applied to our final measurements.

Finally, we compute our measurement errors by using the random point measurements \citep[since the proper noise covariance matrix is not provided, i.e.][]{PlanckCollaboration2015f}, corrected in two ways. First, because we account for the residual CMB primary signal in the $\chi^2$ calculations as discussed below, we remove the error due to the CMB primary itself.   To estimate this contribution, we take 95\% of the minimum covariance between the SPT 150 and 220 GHz bands (filtered to match the \textit{Planck} bands) and the \textit{Planck} 217 GHz band, since these are mostly dominated by the CMB primary signal which will therefore be correlated between them. The minimum covariance is between the two SPT bands, and it is 7.85 $\mu$K. Second, there is an error introduced due to our offset corrections because they are made from a large, but finite number of points. We then get the corrected error from
\begin{equation}
(\sigma_i \sigma_j)_{\text{corr}} = \sqrt{\frac{\sigma_i \sigma_j - \sigma_{\text{cov}}^2}{\text{N}_{\text{source}}} + \frac{\sigma_i \sigma_j}{\text{N}_{\text{random}}}},
\label{eq:newsig}
\end{equation}
where $\sigma_i \sigma_j$ is given by Equation (\ref{eq:sigmaspt}) with $i$ and $j$ representing the various bands used, $\sigma_{\text{cov}} = 7.85 \, \mu$K is the minimum CMB covariance discussed above, N$_{\text{source}}$ is the number of sources used for the measurements (937 for low-$z$ and 240 for high-$z$), and N$_{\text{random}} = 3000$ is the number of random points used.

Note that this represents both the error due to detector noise in each band, as well as the error due to contributions from foregrounds  on the sky.  In fact, the majority of the variance at the highest frequencies is correlated between the bands and likely due to contributions from Galactic dust emission.  However, unlike the primary CMB signal, the spectral shape of this foreground is similar to that of the dusty sources we are trying to constrain, and it cannot be removed by fitting it separately.  

In the same manner as the previous section, we model SPT-SZ 150 and 220 GHz contaminant source fluxes using a range of different source-count models (i.e. $\alpha_d,$ $\alpha_s,$ and $S_{\rm min}$), resulting again in 100,000 modeled contaminating source fluxes in each SPT band, $S_{150,\text{cont}}$ and $S_{220,\text{cont}}$. We also model what the contaminating signal would be in the \textit{Planck} bands and the SPT bands filtered to match \textit{Planck}.  For each modeled contaminating source, if it is chosen to be a synchrotron source we simply extrapolate the \textit{Planck}-based fluxes as
\begin{equation}
S_{\nu,\text{sync}} = S_{150,\text{cont}} \times \left( \frac{\nu}{150} \right)^{\alpha^{150}_{220}} \times C_\nu \times F ,
\label{eq:synchro}
\end{equation}
where $\alpha^{150}_{220}$ is the same used in the previous section, $C_\nu$ is a frequency-dependent factor involved in the conversion from Jy/sr to $\mu$K, and $F = 0.021$ is the factor required to preserve the signal within a 1 arcmin radius aperture after applying the \textit{Planck} filtering we used.

In order to accurately describe  thermal dust emission  across the \textit{Planck} frequencies, we adopt a modified blackbody with a free emissivity index, $\beta$, and dust temperature, $T_\text{dust},$ often referred to as a graybody \citep{PlanckMaps2015}.  This requires us to add another free parameter, the temperature of the contaminant dust, $T_\text{dust}$. 
This slope of each dusty source as a function of frequency is then
\begin{equation}
\left.\frac{\text{d} \ln S_\nu}{\text{d} \ln \nu}\right|_{\nu=185 \text{GHz}} = 3 + \beta - x_{185} [1 - \exp(-x_{185})]^{-1} ,
\label{eq:dustslope}
\end{equation}
where $x_{185} \equiv (185$ GHz)$\times h/(kT) = (185/416) (1+z)/T_{20}$ and $T_{20}$ is the dust temperature in units of 20 K, and
we use the slope of the blackbody function at $\nu = 185$ GHz because it is halfway between our two SPT bands (150 and 220 GHz).  This can be related, in turn, to the power law index $\alpha^{150}_{220},$ from Section \ref{sec:sptcont}, as
\begin{equation}
\beta + 3 = \alpha^{150}_{220} + x_{185} [1 - \exp(-x_{185})]^{-1} .
\label{eq:beta}
\end{equation}
This then gives
\begin{equation}
\begin{split}
S_{\nu,\text{dust}} =& \,\, S_{150,\text{cont}} \times \left( \frac{\nu}{150} \right)^{\alpha^{150}_{220} + x_{185} [1 - \exp(-x_{185})]^{-1}}\\
&\times \frac{\exp[(150/416)(1+z)/T_{20}] - 1}{\exp[(\nu/416)(1+z)/T_{20}] - 1} \\ &\times C_\nu \times F,
\label{eq:sdust}
\end{split}
\end{equation}
where we vary $T_\text{dust}$ from 20 K to 50 K in steps of 3 K.

With these expressions, we are able to compute $\chi^2$ values for each source-count model accounting for the \textit{Planck} measurements. This time, in addition to varying $f_{\text{cont}}$ and $S_{\text{SZ}}$, we also vary T$_\text{dust}$ (as discussed above) and a parameter $\Delta,$ which represents the offset due to the CMB primary signal, which we vary from -3 $\mu$K to 3 $\mu$K in steps of 0.1 $\mu$K. Computing $\chi^2$ now involves the original SPT terms plus the new \textit{Planck} terms, and it follows the same process as in the previous section (e.g. Equation (\ref{eq:chi2matrixes})),
\begin{equation}
\chi^2(f_{\text{cont}},S_{\text{SZ}},T_\text{dust},\Delta) = \mathcal{B} \times \mathcal{A}^{-1} \times \mathcal{B}^T ,
\label{eq:chi2matrixplanck}
\end{equation}
where $\mathcal{B}$ is the signal array and $\mathcal{A}$ is the noise matrix containing the noise for each band plus the covariance terms between each band. We will denote each element of the signal array $\mathcal{B}_i$, where $i$ runs over the two SPT bands (i.e. 150 and 220 GHz) and then every \textit{Planck}-filtered band (i.e. the \textit{Planck} bands at 857, 545, 353, and 217 GHz, plus the SPT bands at 220 and 150 GHz filtered to match the \textit{Planck} images). We then have $\mathcal{B}_1 = f_{\text{cont}} \times \left<S_{150,\text{cont}}\right> + S_{\text{SZ}} - S_{150}$, $\mathcal{B}_2 = f_{\text{cont}} \times \left<S_{220,\text{cont}}\right> - S_{220}$, and $\mathcal{B}_{3-8} = f_{\text{cont}} \times \left<S_{3-8,\text{cont}}\right> + \Delta - S_{3-8}$. As before, $S_i$ represents the final values of our galaxy stacks for each band.  We similarly define the elements of the noise matrix as $\mathcal{A}_{i j} = \sigma_i \sigma_j$, where $i$ and $j$ run over all of the bands and $\sigma_i \sigma_j$ is given by Equation (\ref{eq:newsig}).

As in the previous section, we then convert the $\chi^2$ values to Gaussian probabilities by taking
\begin{equation}
\begin{split}
& P(S_{\text{SZ}}) = \\ & \sum_{f_{\text{cont}} \in [0,1],T_\text{dust},\Delta} \frac{\exp[-\chi^2(f_{\text{cont}},S_{\text{SZ}},T_\text{dust},\Delta)/2]}{\sum_{\text{all}}\exp[-\chi^2/2]},
\label{eq:probplanck}
\end{split}
\end{equation}
where the whole function is normalized to a total of 1, and each final SZ value contains the sum over the corresponding $T_\text{dust}$, $\Delta$, and fractions from 0 to 1. Since in this case there are  8 terms contributing to $\chi^2$ and 4 fit parameters, this leaves us with 4 degrees of freedom.
Thus the minimum $\chi^2$ will not be 0 in every case as they were previously with just 2 parameters and 2 fit parameters, and so for each model we scale the final probabilities by $\exp(-\chi^2_{\text{min}}/2)$, where $\chi^2_{\text{min}}$ is the minimum $\chi^2$ value for that model.

\begin{figure*}[t]
\centerline{\includegraphics[height=10cm]{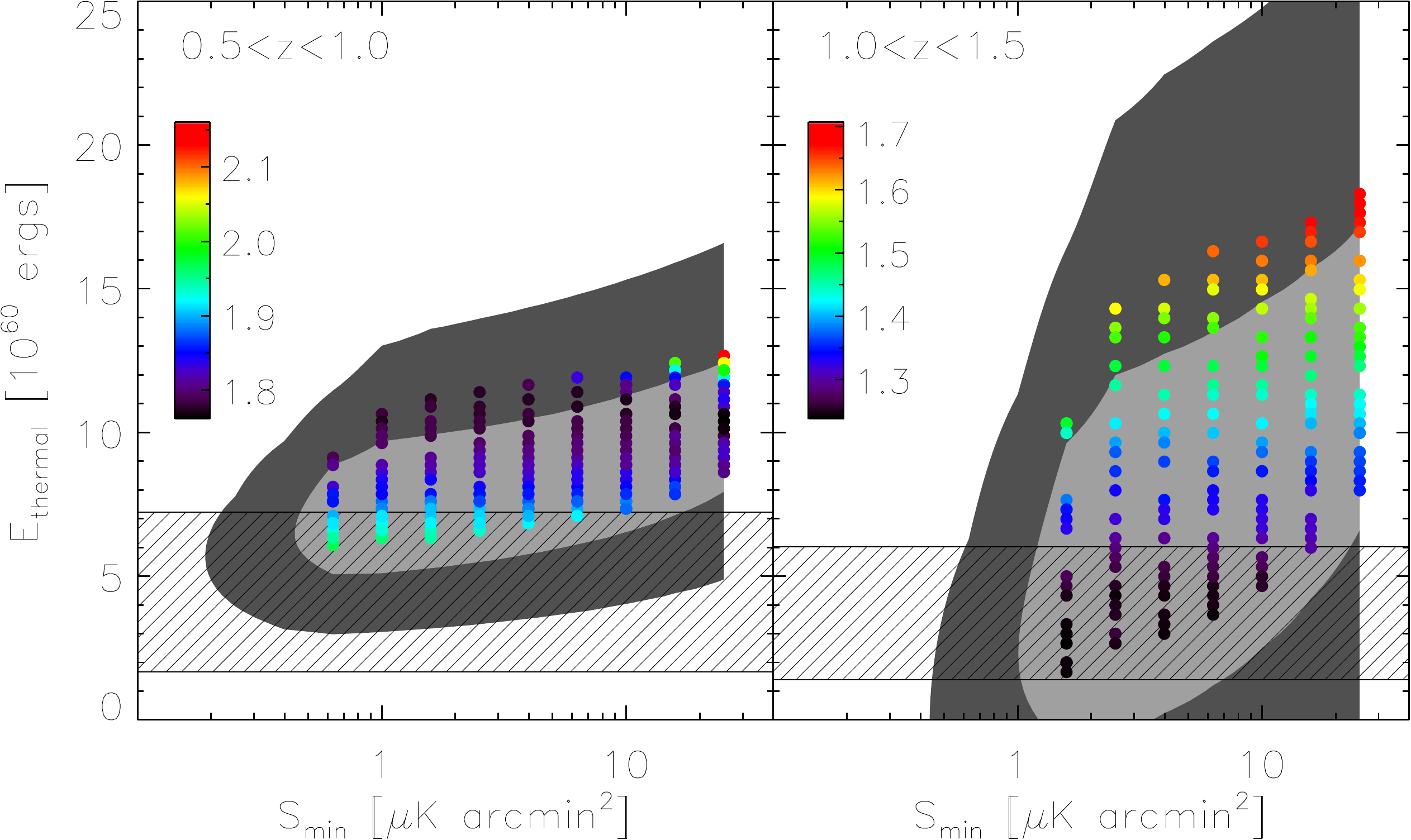}}
\caption{\small Plot of the contaminant-corrected $E_\text{therm}$ (see Equation (\ref{eq:newEthrm})) for different choices of $\alpha_{\rm dust}$, $\alpha_{\rm sync}$, and $S_{\rm min}$, incorporating the \textit{Planck} bands. Points are located at the peak $\chi^2$ probability for each model, and colored according to the minimum $\chi^2$ value for that model. Their general locations are still indicative of their $\alpha_d,$ $\alpha_s,$ and $S_{\rm min}$ values, as seen in Figure \ref{fig:s150corr}. The light and dark gray regions represent the complete span of $\pm1\sigma$ and $\pm2\sigma$, respectively, for all points. The hatched regions represent the $\pm1\sigma$ range for $E_\text{grav}$ (see Equation (\ref{eq:Egrav})). \vspace{3mm}}
\label{fig:s150corrPlanck}
\end{figure*}

This again gives us a function $P(S_{\text{SZ}})$ for each combination of $\alpha_d,$ $\alpha_s,$ and $S_{\rm min}$, which we can convert to an energy $E_\text{therm}$. The peak of each $P(S_{\text{SZ}})$ curve is shown as the colored points in Figure \ref{fig:s150corrPlanck}, where the points are colored by the minimum $\chi^2$ value for each model.  Note that the the best fit $\chi^2$ values are smaller than expected for 4 degrees of freedom due to the correlations between the errors of the various frequency bands due primarily to foreground contamination by Galactic dust.
The 1$\sigma$ and 2$\sigma$ contours are created for each $S_{\rm min}$ by averaging $P(S_{\text{SZ}})$ across $\alpha_d$ and $\alpha_s$ and then dividing the final result by the single maximum value. 1$\sigma$ and 2$\sigma$ are again represented by the values 0.61 and 0.13, respectively. These contours are shown in Figure \ref{fig:s150corrPlanck}, along with the $\pm1\sigma$ range for $E_\text{grav}$. From this figure we can see that the $\sigma$ values have slightly decreased at both low-$z$ and high-$z$, especially at higher $S_{\rm min}$ values for high-$z$. This is where the contaminants are hardest to constrain with just SPT, and where \textit{Planck} data helps us the most.

Finally, we average the probability distribution across $S_{\rm min}$, divided by the maximum value again, and get a final distribution as a function of only $E_\text{therm}$. The significance ($\sigma$) values of this curve are shown in the ``With Planck'' part of Table \ref{tab:finvals}. \textit{Planck} has helped to constrain the tSZ signal, especially at high-$z$, although it is clear that the gain in sensitivity has been limited by the decrease in the number of galaxies in each redshift bin due to the much larger beam size of \textit{Planck} compared to SPT.

Alternatively, we can also characterize the total tSZ signal for our coadds with the angularly integrated Compton-$y$ parameter, $Y$ \citep[e.g.,][]{Ruan2015}. While we cannot directly compare peak Compton-$y$ values with past measurements, as these are beam-dependent quantities, we can compare the angularly integrated $Y$ values between our results and past experiments (see Table \ref{tab:prevwork}). Using Equation (\ref{eq:newys}) at 150 GHz, this is
\begin{equation}
\begin{split}
Y &\equiv l_{\rm ang}^2 \int y(\bm{\theta}) d\bm{\theta} \\ &= -3.5 \times 10^{-8} \, \text{Mpc}^2 
 \left(\frac{l_{\rm ang}}{\text{Gpc}}\right)^2 \frac{\int \Delta T_{150}(\bm{\theta}) d\bm{\theta}}{\text{$\mu$K arcmin$^2$}},
\label{eq:bigy}
\end{split}
\end{equation}
such that $Y = 2.9 \times 10^{-8}  \, \text{Mpc}^2 E_{\rm 60}$, where $E_{\rm 60}$ is $E_\text{therm}$ in units of $10^{60}$ erg.
In these units, the mean $Y$ values computed directly from the 150 GHz maps from our co-added galaxies are $1.2 (\pm 0.6) \times 10^{-7}$ Mpc$^2$ for low-$z$ and $-1.7 (\pm 1.5) \times 10^{-7}$ Mpc$^2$ for high-$z.$
When these measurements are corrected for contamination using the $220$ GHz SPT data, the mean $Y$ values become $2.3_{-0.7}^{+0.9}\times 10^{-7}$ Mpc$^2$ for low-$z$ and $1.9_{-2.0}^{+2.4}\times 10^{-7}$ Mpc$^2$ for high-$z,$ and once the Planck data is incorporated, the mean $Y$ values become $2.2_ {-0.7}^{+0.9}  \times 10^{-7}$ Mpc$^2$ for low-$z$ and $1.7_ {-1.8}^{+2.2} \times 10^{-7}$ Mpc$^2$ for high-$z.$ These values are also given in Table \ref{tab:finvals}.

\begin{table*}[t]
\begin{center}
\resizebox{16cm}{!}{
\begin{tabular}{|c|c|c|c|c|c|c|}
\hline
Study                           & N            & Type             & $z$   & Mass ($M_\odot$)           & tSZ Measurement                          & Type \\ \hline
\citet{Chatterjee2009} & 500,000 & SDSS quasars & $0.08-2.82$ & $-$ & (7.0$\pm$3.4) $\times 10^{-7}$ & $y$ \\              
\citet{Chatterjee2009} & 1,000,000 & SDSS LRGs & $0.4-0.6$ & $-$ & (5.3$\pm$2.5) $\times 10^{-7}$ & $y$ \\
\citet{Hand2011}         & 1732         & SDSS radio-quiet LRGs & 0.30 (mean) & 8.0 $\times 10^{13}$ & (7.9$\pm$6.2) $\times 10^{-7}$ Mpc$^2$ & $Y_{200\bar{\rho}}$ \\
\citet{Gralla2014} & 667 & SDSS radio-loud AGN & $0.3$ (median) & $2 \times 10^{13}$ & (1.5$\pm$0.5) $\times 10^{-7}$ Mpc$^2$ & $Y_{200}$ \\
\citet{Gralla2014} & 4,352 & FIRST AGN & $1.06$ (median) & $-$ & (5.7$\pm$1.3) $\times 10^{-8}$ Mpc$^2$ & $Y_{200}$ \\
\citet{Greco2014} & 188,042 & SDSS LBGs & $0.05-0.3$  & $1.4 \times 10^{11} \star$  & $(0.6_{-0.6}^{+5.4})$ $\times 10^{-6}$ arcmin$^2$ & $\tilde{Y}^{cyl}_c$ \\
\citet{Ruan2015} & $\approx$14,000 & SDSS quasars & $1.96$ (median)  & $5.0 \times 10^{12}$ & (4.8$\pm$0.8) $\times 10^{-6}$ Mpc$^2$ & $Y$ \\
\citet{Ruan2015} & $\approx$14,000 & SDSS quasars & $0.96$ (median)  & $5.0 \times 10^{12}$ & (2.2$\pm$0.9) $\times 10^{-6}$ Mpc$^2$ & $Y$ \\
\citet{Ruan2015} & 81,766 & SDSS LBGs & $0.54$ (median)  & $3.2 \times 10^{11} \star$ & (1.4$\pm$0.4) $\times 10^{-6}$ Mpc$^2$ & $Y$ \\
\citet{Crichton2015} & 17,468 & SDSS radio-quiet quasars & $0.5-3.5$ & $-$ & (6.2$\pm$1.7) $\times 10^{60}$ erg & $E_{th}$ \\
\hline
\end{tabular} }
\end{center}
\caption{\small Previous tSZ measurements. LRGs = luminous red galaxies; LBGs = locally brightest galaxies. Masses refer to halo masses, except for those of \citet{Greco2014} and \citet{Ruan2015} LBGs which refer to stellar masses ($\star$). We select \citet{Hand2011} and \citet{Greco2014} values that have the most similar masses to our galaxies.}
\label{tab:prevwork}
\end{table*}

\begin{table*}[t]
\begin{center}
\resizebox{15cm}{!}{
\begin{tabular}{|c|c|c|c|c|c|c|c|}
\hline
Model & N & $z$ & $\int \Delta T_{150}(\bm{\theta}) d\bm{\theta}$ & $Y$ & $E_\text{therm} (\pm1\sigma)$ & $E_\text{therm} (\pm2\sigma)$ & S/N \\ 
          &    &        &  ($\mu$K arcmin$^2$)                                         & ($10^{-7}$ Mpc$^2$) & ($10^{60}$ erg) & ($10^{60}$ erg) & ($S/\sigma$)  \\  \hline
Data only & 3394  & $0.5-1.0$ & $-1.5 \pm 0.7$  & $1.2 \pm 0.6$ & $4.1 \pm 1.9$ & $ 4.1 \pm 3.8$ & 2.2  \\              
 & 924  & $1.0-1.5$ & $1.6 \pm 1.4$  & $-1.7 \pm 1.5$ & $-5.8 \pm 5.1$ & $-5.8\pm 10.2$ & -1.1  \\
$\chi^2$ (SPT only) & 3394 & $0.5-1.0$ & $-2.9_{-1.1}^{+0.9}$  & $2.3_{-0.7}^{+0.9}$ & $8.1_{-2.5}^{+3.0}$ & $8.1_{-4.8}^{+6.8}$ & 3.5  \\
 & 924 & $1.0-1.5$ & $-1.8_{-2.3}^{+1.9}$  & $1.9_{-2.0}^{+2.4}$ & $6.7_{-7.0}^{+8.3}$ & $6.7_{-13.3}^{+18.6}$ & 0.9  \\
$\chi^2$ (With Planck) & 937 & $0.5-1.0$ & $-2.8_{-1.1}^{+0.8}$  & $2.2_ {-0.7}^{+0.9}  $ & $7.6_{-2.3}^{+3.0}$ & $7.6_{-4.3}^{+7.1}$ & 3.6  \\
 & 240 & $1.0-1.5$ & $-1.7_{-2.1}^{+1.7}$  & $1.7_ {-1.8}^{+2.2}$ & $6.0_{-6.3}^{+7.7}$ & $6.0_{-12.3}^{+18.0}$ & 0.9  \\
\hline
\end{tabular} }
\end{center}
\caption{\small Our final tSZ measurements using various methods for removing contamination. The last three columns represent the best fit $E_\text{therm}$ values with $\pm1\sigma$ values and $\pm2\sigma$ values and the $E_\text{therm}$ signal-to-noise ratio ($S$/$\sigma$), respectively.}
\label{tab:finvals}
\end{table*}

Our $0.5 \leq z \leq 1.0$ value of $2.2_ {-0.7}^{+0.9}  \times 10^{-7}$ Mpc$^2$ is more than 3 times smaller than the \citet{Hand2011} $z\approx 0.3$ SDSS radio-quiet LRG result. If we estimate the stellar mass for the \citet{Hand2011} results using Table \ref{tab:prevwork} and Equation (\ref{eq:Mstellar}) we get $1.7\times10^{12} M_\odot$, which is about an order of magnitude greater than the average stellar mass of our galaxies. Our smaller values could be indicative of the relation that tSZ signal increases with halo (and stellar) mass \citep[i.e.,][]{Gralla2014}. Our low-$z$ result is within about 1$\sigma$ of the \citet{Gralla2014} $z\approx 0.3$  SDSS radio-loud AGN result, though our $1.0 < z \leq 1.5$ result of $1.7_ {-1.8}^{+2.2} \times 10^{-7}$ Mpc$^2$ is $> 3\sigma$ higher than their $z\approx 1.1$ FIRST AGN result. This discrepancy is not too significant because our high-$z$ detection is only at a 0.9$\sigma$ confidence. Our results are also within $\approx$$1\sigma$ of \citet{Greco2014} when comparing their results for galaxies with masses similar to ours. At smaller masses our results are consistent with theirs while at larger masses they find even greater tSZ signal, following the relation that tSZ signal increases with stellar mass.  \citet{Ruan2015} also obtain values from stacks of SDSS quasars about an order of magnitude ($> 2\sigma$) higher than ours, although, according to \citet{Cen2015}, the maximum AGN feedback signal from \citet{Ruan2015} can only be 25\% of their quoted values. Furthermore, the  $\approx 10^{11} M_\odot$ galaxy results from \citet{Ruan2015} are consistent with zero signal, while their $\approx 3 \times 10^{11} M_\odot$ galaxy results are $>2\sigma$ larger than ours. Their high mass sample represents almost 3 times the mean mass of our galaxies, though, so the larger values may be indicative of the stellar mass$-$tSZ signal relation  as well as the potential overestimation of the tSZ signal. We can compare our results to \citet{Cen2015} by multiplying their average Compton-$y$ values over 1 arcmin by $\pi \times l_{\text{ang}}^2$ (where $l_{\text{ang}}^2 = 3.18 \,\text{Gpc}^2$ for $z=1.5$)  to get $Y$ values of $\approx$ 6.8$\times 10^{-7}$ Mpc$^2$ for their halo occupation distribution (HOD) model, and $\approx$ 4.2$\times 10^{-7}$ Mpc$^2$ for their \citet[][CS]{Cen2015a} model, which places quasars in lower mass dark matter halos. This CS model value is within $\approx$ 2$\sigma$ of our results, and our measurements would favor the lower estimates of their CS model over their HOD model. Finally, both our low-$z$ and high-$z$ $E_\text{therm}$ results are well within $1\sigma$ of the \citet{Crichton2015} SDSS radio-quiet quasar results.

With Equations (\ref{eq:Egrav}) and (\ref{eq:EAGN}) and the redshifts and masses from Table \ref{tab:meanvals}, we can also investigate theoretical thermal energies of the gas around elliptical galaxies due to both gravity and AGN feedback.  We estimate the gravitational heating energy to be $E_\text{therm,grav} = 3.6_{-1.9}^{+3.6} \times 10^{60}$ erg for our low-$z$ sample and $E_\text{therm,grav} = 3.0_{-1.6}^{+3.0}\times 10^{60}$ erg for our high-$z$ sample. We therefore measure excess non-gravitational energies (for our results using \textit{Planck}) of $E_\text{therm,feed,dat} = 4.0^{+3.6}_{-4.3} \times 10^{60}$ erg for low-$z$ and $E_\text{therm,feed,dat} = 3.0^{+7.9}_{-7.0} \times 10^{60}$ erg for high-$z$. Plugging these into Equation (\ref{eq:EAGN}) and solving for $\epsilon_{k}$, we get feedback efficiencies of $7.5^{+6.5}_{-8.0}$\% for low-$z$ and $6.5^{+17.5}_{-15.5}$\% for high-$z$. These values are very uncertain, though they are consistent with the suggested 5\% \citep[i.e.,][]{Scannapieco2005,Ruan2015}.

\subsection{Anderson-Darling Goodness-of-fit Test}
\label{sec:ad}

The measurements described above depend on co-adding data from a large number of sources.   In principle, however, there is additional information in the distribution of measured values that is lost through this process.   For example, imagine a set of 1001 150 GHz measurements, 1000 of which contributed a negative signal of -2 $\mu K$ arcmin$^2$ and one of which contributed a positive signal of 2000 $\mu K$ arcmin$^2$.  While the average value of these measurements would be zero, looking at the distribution of values would indicate strong evidence of a negative tSZ signal, offset by contamination from a single, overpowering positive source.   

To quantify the additional information available by the full distribution of SPT data, we apply the same contaminant source modeling described above (i.e. Section \ref{sec:sptcont}) and use a goodness-of-fit test, the Anderson-Darling (A-D) test \citep{Anderson1954}, to find models that poorly fit the data.  In this case we restrict our attention purely to the $1$ arcmin resolution data used to construct Figure \ref{fig:coaddims}, without folding in the lower resolution \textit{Planck} data as described in Section \ref{sec:planckcont}.  To perform the test, we run through every pair of galaxy measurements (i.e., the 1 arcmin radius aperture sums at both 150 and 220 GHz) and find the fraction of galaxy measurements in each of the four quadrants around the pair of measurements (i.e., ($x<x_i, y<y_i$); ($x<x_i, y>y_i$); ($x>x_i, y<y_i$); ($x>x_i, y>y_i$); where $x$ and $y$ represent the co-add sum in each band, and $i$ runs through all the galaxies). We will call these fractions $f_{i,j}$ where $i$ specifies the galaxy ($i = [1,2,\dots,n]$, with $n$ being the number of galaxies) and $j$ specifies the quadrant ($j = [1,2,3,4]$). In the same four quadrants we also find the fraction of model measurements, $F_{i,j}$. We therefore define our A-D statistic as:
\begin{equation} S_{AD} = n \, \sum \frac{(f_{i,j}-F_{i,j})^2}{F_{i,j}(1-F_{i,j})} ,\end{equation}
where a smaller $S_{AD}$ indicates a better fit between the model and the data.

\begin{figure*}[t]
\centerline{\includegraphics[height=8cm]{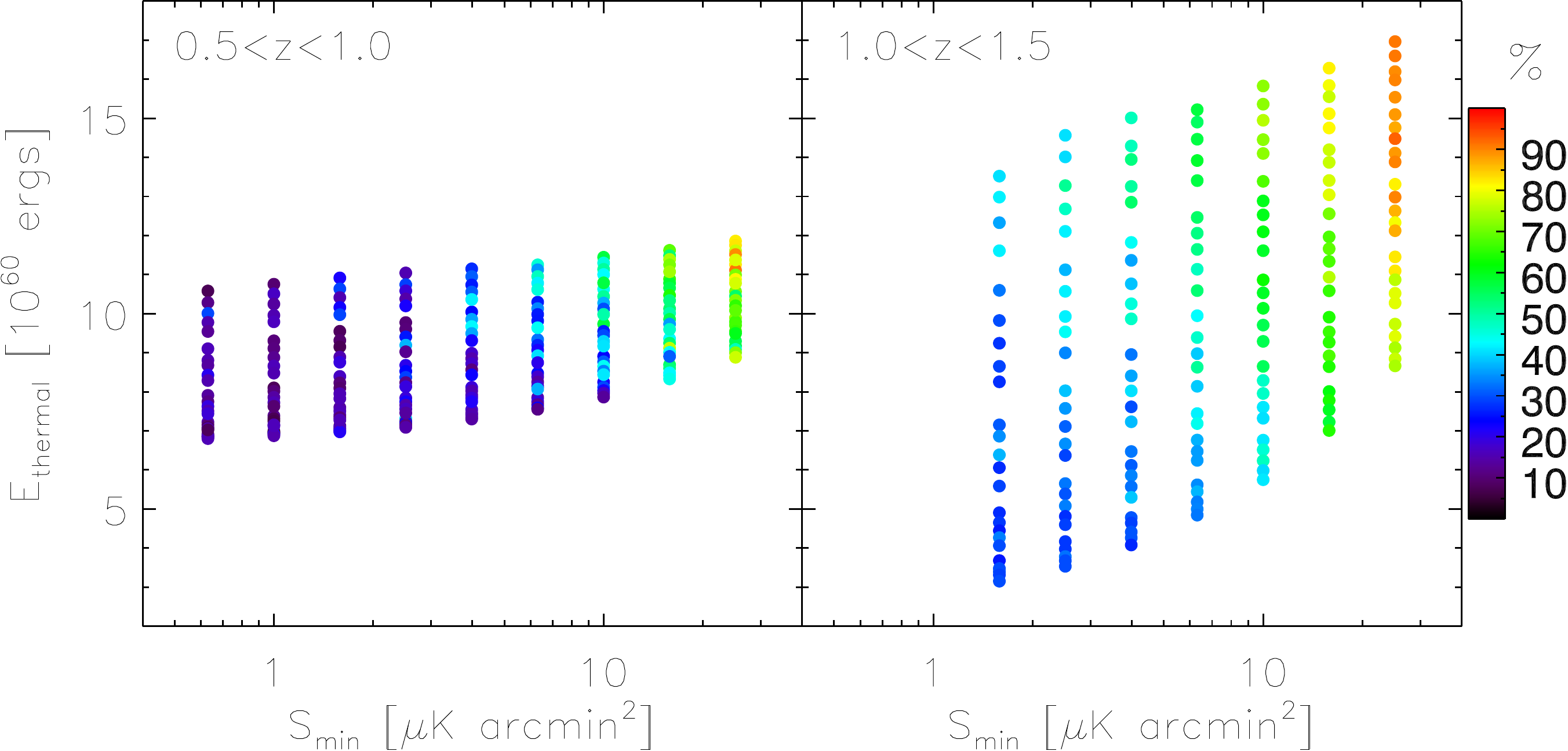}}
\caption{\small Same data points as Figure \ref{fig:s150corr}, but with the points colored according to the A-D statistics. \vspace{3mm}}
\label{fig:testfigs}
\end{figure*}

Each model, as a function of $S_{\rm min}$, $\alpha_{\rm dust}$, and $\alpha_{\rm sync}$, will have a corresponding $S_{AD}$. In order to interpret our results, we follow the same process, but instead of using data from our selected galaxies we use a random subset of the modeled sources, with the subset containing the same number of elements as the number of galaxies we are using for both redshift ranges. We do this 200 times each for four different combinations of $S_{\rm min}$, $\alpha_{\rm dust}$, and $\alpha_{\rm sync}$. We then define our confidence of the model fits, $P(<S_{AD})$, as the fraction of these subset calculations that are less than the corresponding $S_{AD}$. When defined in this way, $P$ indicates our confidence that the model is not a good fit with the data. The fractions correspond to $\sigma$ in the standard way (i.e., 1$\sigma$ = 0.68, 2$\sigma$ = 0.95, etc.).

As mentioned above, we only do these subset calculations for four sets of model parameters, which is due to the time-intensive nature of these computations. To get confidence values for every other set of model parameters, we simply do a linear interpolation between the four sets of model parameters we did use. The results are shown in Figure \ref{fig:testfigs}. They reveal that models with both the highest $S_{\text{min}}$ and the highest $\alpha_{\rm dust}$ are disfavored up to $\approx 1.5\sigma$ (87\%) confidence, with the trend much more pronounced in the high-$z$ range.  On the other hand, unlike the analysis using the \textit{Planck} data, the A-D test primarily serves to constraint $S_{\text {min}}$ rather than $E_{\rm thermal},$ meaning that it does not allow us to obtain significantly better constraints on feedback itself.  Finally, we note that we also carried out a two-dimensional Kolmogorov-Smirnov test \citep[e.g.,][]{Peacock1983} but it was far less constraining than the A-D test, and so we do not present it here.


\section{Discussion}

Since $z \approx 2,$ star formation has occurred in progressively less massive galaxies, and AGNs have occurred around progressively less massive black holes.  While these are fundamental observations of galaxy evolution,  a consensus has yet to be reached about the physical processes that dictate them.  The anti-hierarchical quenching of galaxies and AGNs might be partially caused by stable virial shocks and gravitational heating due to infalling galaxies \citep[e.g.,][]{Feldmann2015}, but most successful models invoke additional energy input, most likely from AGNs.   In fact, strong quasar activity is known to launch rapid outflows of gas, and powerful radio jets are observed to play an important role in galaxy clusters, but the total energy released by these processes as a function of redshift and environment remains largely unknown. As our understanding of galaxy formation increasingly relies on  understanding this feedback, it is apparent that we need increasingly sensitive observations to constrain it.

An extremely promising approach to making these constraints is co-adding the microwave background around a large number of sources to measure the signal imprinted by the tSZ effect.  Several recent studies have applied this approach, making detections of galaxies at low redshifts ($z \lesssim 0.5$)  and AGNs from $z=0$ to 3, as summarized in Table \ref{tab:prevwork}. There are potential issues with each of these approaches, though. At low redshifts the additional gravitational heating from structure formation obscures the additional energy input from AGNs, while working with AGNs directly leads to problems of strong contamination from dust and synchrotron emission.  These have motivated us to choose massive ($> 10^{11} M_\odot$) elliptical galaxies at moderate redshifts ($0.5 \leq z \leq 1.5$), where we expect these various limitations on the AGN feedback signal to be minimal, and to make our measurements using data from the South Pole Telescope, which has a $\approx$ 1 arcmin beam size well matched to the expected sizes of heated regions.

To construct a catalog of such large, $0.5 \leq z \leq 1.5$ elliptical galaxies, we made use of data from the  BCS in the $g$, $r$, $i$, and $z$  bands, as well as  VHS  data in the $J$, $H$, and  $K_s$ bands over a $\approx 43$ deg$^2$ area overlapping with the public SPT fields.  We separated galaxies from stars using a $gzK_s$ color cut, and for each of the galaxies, we fit stellar population synthesis models to limit the sample to the most massive, $z \geq 0.5$ passive galaxies.   Furthermore, to limit the contamination of the tSZ signal, we removed all galaxies if they were within 4 arcmin of a galaxy cluster, an active AGN, a dusty Galactic molecular cloud, or a galaxy with strong dust emission.  Finally, around the remaining sources, we co-added 150 and 220 GHz SPT maps that were optimally filtered for point sources. This alone gave us a tSZ detection in our low-$z$ subset ($0.5 \leq z \leq 1.0$) 150 GHz band of $> 2\sigma$ significance. At the same time, we also had a $> 2\sigma$ contaminant signal in our low-$z$ 220 GHz band, which is expected to also extend to and contaminate the 150 GHz band.
 
In order to account for this contamination, we modeled the potential contaminating sources using the SPT point source number counts from \citet{Mocanu2013}, extrapolated them between the two bands, and used $\chi^2$ statistics to get best-fit values across all reasonable parameter choices. This improved our low-$z$ subset tSZ detection to $3.5\sigma$ significance. To even further constrain the contamination in our measurements, we stacked our galaxies in the four highest \textit{Planck} bands as well, rejecting galaxies within 10 arcmin of potential contaminant sources. We again used $\chi^2$ statistics to get best-fit values across all reasonable parameter choices, and found a low-$z$ subset tSZ detection at 3.6$\sigma$ significance, as well as a $0.9 \sigma$ measurement of the tSZ signal in the high-$z$ subset ($1.0 \leq z \leq 1.5$). A summary of all our results is found in Table \ref{tab:finvals}.

In comparison with previous work measuring the tSZ signal around AGNs, we find a similar and slightly larger ($\approx 1\sigma$) tSZ signal than the lower redshift results of \citet{Gralla2014}, and a significantly larger ($> 3\sigma$) signal than their higher redshift results. However, we find a much smaller  tSZ signal than both the high and low redshift results of \citet{Ruan2015}, although it is suspected that their values are significantly overestimated \citep[e.g.,][]{Cen2015}. We also find a slightly smaller tSZ signal than the simulated results of \citet{Cen2015}, and our results favor their CS model, which associates quasars with lower mass dark matter halos. Our $E_\text{therm}$ results are consistent with \citet{Crichton2015}. In comparison with previous work measuring the tSZ signal around galaxies, we find a signal that is about 3.5 times less than the more massive galaxies used by \citet{Hand2011} which may be indicative of the stellar mass$-$tSZ signal relation. We find a signal within $\approx1\sigma$ of the similar-mass galaxy stacks of \citet{Greco2014}, though our results are consistent with their results at lower masses. The lower mass \citet{Ruan2015} galaxy signal is consistent with zero, while the higher mass ($\approx 3$ times our mean mass) results are $>2\sigma$ larger than ours. This may again reflect the mass$-$tSZ signal relation as well as their potential overestimation of the tSZ signal. Finally, our low redshift results suggest an AGN feedback efficiency of $7.5^{+6.5}_{-8.0}$\%, which is consistent with the 5\% value found in \citet{Ruan2015} and suggested by, for example, \citet{Scannapieco2005}.

Measurements such as the one described here are likely to improve significantly in the near future. While the  first public SPT-SZ data release (2011) covers a 95 deg$^2$ field with the 150 and 220 GHz bands, the upcoming full survey release will include a 2500 deg$^2$ field using bands at 95, 150, and 220 GHz. The much larger field will allow for a much larger set of galaxies to be co-added, vastly improving the signal-to-noise of the measurements, while the additional 95 GHz band will also allow for further constraints on contaminating signals.  In addition to SPT, the Atacama Cosmology Telescope (ACT) has observed for four seasons from 2007 to 2011 using the Millimeter Bolometric Array Camera (MBAC) with bands at 148, 218, and 277 GHz, producing more than 90 TB of data \citep{Dunner2013}. In 2012 they released a 780 deg$^2$ temperature map at 148 GHz\footnote{\scalebox{0.8}{http://lambda.gsfc.nasa.gov/product/act/act\_tmaps\_info.cfm}}, in 2014 they released a few thousand deg$^2$ at 148 and 218 GHz\footnote{\scalebox{0.8}{http://lambda.gsfc.nasa.gov/product/act/act\_maps2013\_info.cfm}}, and more fields using all 3 bands will be released in the future. Measuring galaxies using ACT can compliment work using SPT because they observe both different and overlapping regions of the sky. Furthermore, the higher frequency 277 GHz band can also provide important help in constraining contaminant signal.  In the future, separating out such contaminants will become even more practical, through surveys such as those to be undertaken by the upgraded ACT telescope (Advanced ACTPol) and the proposed Cerro Chajnantor Atacama Telescope (CCAT).\footnote{\scalebox{0.8}{http://www.ccatobservatory.org/}}

Another approach to  constraining AGN feedback is through deep measurements of smaller samples of galaxies identified as the most interesting using large radio telescopes.  In this case rather than co-adding as many galaxies as possible, one would select a handful of the most promising galaxies for detecting AGN feedback. The Goddard IRAM Superconducting Two Millimeter Camera (GISMO)  and the New IRAM KIDs Array (NIKA) are powerful new instruments mounted on the Institute de Radioastronome Millimetrique (IRAM) 30 meter telescope\footnote{\scalebox{0.8}{http://www.iram-institute.org/EN/30-meter-telescope.php}} that may prove useful for this purpose. Also promising is the National Radio Astronomy Observatories (NRAO) Green Bank Telescope (GBT),  whose Continuum Backend operates at lower frequencies where the tSZ signal is roughly three times larger.
On the other hand, interferometers appear to be less suited to constraining AGN heating, because they are more likely to resolve the affected regions and thus be limited by surface brightness concerns. Nevertheless, several interferometers may prove useful for AGN feedback studies, including the IRAM interferometer, the Combined Array for Research in Millimeter-wave Astronomy (CARMA),  and the Atacama Large Millimeter/sub-millimeter Array (ALMA).

Finally, tSZ simulations and observations can be combined to produce weighted stacks that are adapted to be as sensitive as possible to the differences between feedback models.   This is because with a suite of simulations in hand, one can not only perform stacks of the tSZ signal around simulated galaxies with exactly the same mass, redshift, and age distribution as in a given observational sample, but also vary the weights applied to such stacks so as to arrive at the combination that allows for the observations to best discriminate between competing models.    We are only now beginning to map out the history of AGN feedback through measurements of the tSZ  effect.


We would like to thank Shantanu Desai, Mariska Kriek, Eliot Quataert, and Daniel Marrone for helpful discussions. We would also like to thank the anonymous referee for their very helpful comments. This work was supported by the National Science Foundation under grant AST14-07835. ES gratefully acknowledges Joanne Cohn, Eliot Quataert, the UC Berkeley Theoretical Astronomy Center, and Uro\v{s} Seljak and the Lawrence Berkeley National Lab Cosmology group, for hosting him during the period when a substantial portion of this work was carried out.

\bibliographystyle{apj}
\small
\bibliography{references}
\end{document}